# Revisiting the luminescence properties of $Pr^{3+}$ : YAG within the framework of an extended approach of Judd-Ofelt theory

M. Lepers[1*], G. Hovhannesyan[2], Y. Guyot[3], R. Moncorgé[4*], M. Velazquez[5]

[1] *Université Bourgogne Europe, CNRS, Laboratoire Interdisciplinaire Carnot de Bourgogne, ICB UMR 6303, 21000 Dijon, France*

[2] *Université Paris-Saclay, CNRS, Laboratoire Aimé Cotton, Orsay 91400, France*

[3] *Université de Lyon, CNRS, Institut Lumière Matière, 10 rue Ada Byron, 69622 Villeurbanne, France*

[4] *Université de Caen Normandie, Centre de recherche sur les Ions, les Matériaux et la Photonique, UMR 6252 CNRS-CEA-ENSICAEN-UNICAEN, 14050 Caen, France*

[5] *Université de Grenoble Alpes, Laboratoire SIMAP, UMR 5266 CNRS-UGA-Grenoble INP, 38420 Saint Martin d'Hères, France*

*Corresponding authors : maxence.lepers@u-bourgogne.fr, richard.moncorge@ensicaen.fr

**Keywords :** Luminescence, Praseodymium, YAG, ZBLAN, Extended Judd-Ofelt theory

**Abstract :**

We show in this article the improvements which can be obtained in the description of the luminescence properties of $Pr^{3+}$ doped materials by using an extension of the Judd-Ofelt theory in order to relax some strong selection rules and approximations of the standard formalism and to better account for the influence of the 4f5d excited electronic configuration. The demonstration is made by re-examining the case of $Pr^{3+}$:YAG, a well known luminescent and laser crystal with a very low energy 4f5d absorption band. Our extension thus provides a better agreement between calculated and measured absorption intensities, especially for the hypersensitive $^3H_4 \rightarrow {}^3P_2$ transition. A comparison is made with the results obtained in the case of $Pr^{3+}$:ZBLAN, a laser fluoride glass with much higher 4f5d absorption levels. Our investigation also gives the opportunity, in the case of $Pr^{3+}$:YAG, to provide more complete and more reliable absorption and emission data than reported in the past literature and to exploit these data to better address the question of laser operation at various emission wavelengths. It is thus demonstrated that laser operation should be possible with improved laser performance at 488 nm, 616 nm and 744 nm, as it was already achieved in the past, but also at 566 nm and 931 nm by using appropriate laser cavities and laser mirrors.

# Revisiting the luminescence properties of Pr$^{3+}$ : YAG within the framework of an extended approach of Judd-Ofelt theory

M. Lepers[1*], G. Hovhannesyan[2], Y. Guyot[3], R. Moncorgé[4*], M. Velazquez[5]

[1] Université Bourgogne Europe, CNRS, Laboratoire Interdisciplinaire Carnot de Bourgogne,
ICB UMR 6303, 21000 Dijon, France

[2] Université Paris-Saclay, CNRS, Laboratoire Aimé Cotton, Orsay 91400, France

[3] Université de Lyon, CNRS, Institut Lumière Matière, 10 rue Ada Byron, 69622 Villeurbanne, France

[4] Université de Caen Normandie, Centre de recherche sur les Ions, les Matériaux et la Photonique,
UMR 6252 CNRS-CEA-ENSICAEN-UNICAEN, 14050 Caen, France

[5] Université de Grenoble Alpes, Laboratoire SIMAP, UMR 5266 CNRS-UGA-Grenoble INP,
38420 Saint Martin d'Hères, France

*Corresponding authors : maxence.lepers@u-bourgogne.fr, richard.moncorge@ensicaen.fr

## 1. Introduction

Optical transitions between energy levels of the 4f$^2$ ground-state electronic configuration of Pr$^{3+}$ ions are always strongly influenced by the proximity of the 4f5d first-excited one (see in [1] and [2], for instance). This is the reason why, in addition to some strong selection rules and approximations, the standard Judd-Ofelt (J.O.) formalism [3, 4] often fails to give correct descriptions of measured absorption intensities and reliable predictions of radiative emission lifetimes and branching ratios (see in [5-7], for instance], and why numerous works were dedicated in the past at the modification and the improvement of this formalism. In the case of Pr$^{3+}$, this includes the method hereafter designated as « Modified » or KM (for Kornienko) method consisting in the introduction [8-9] of a fourth adjustable parameter α (in addition to the usual $\Omega_2$, $\Omega_4$ and $\Omega_6$ J.O. parameters) which should be (in principle) inversely proportional to some 4f5d energy, and the method consisting in the introduction of supplementary odd rank third order intensity parameters $\Omega_1$, $\Omega_3$ and $\Omega_5$ [10], the ajustments of the experimental data being finally obtained with the aid of at least five adjustable parameters.

Our purpose here is to show the improvements which can be obtained by using an extension of the theory allowing to relax some of the strong assumptions and selection rules of the Judd-Ofelt theory, and which was recently applied successfully to the case of Eu$^{3+}$, Nd$^{3+}$ and Er$^{3+}$ doped materials [11, 12]. Such a extension allows to relax some of the strong assumptions of the J.O. theory, for instance the strict application of the closure relation ; it also allows the adjustment of the experimental data with only three odd rank third order (strictly positive) intensity parameters labelled $X_1$, $X_3$ and $X_5$, the coefficients being calculated by using the free-ion wavefunctions computed using a Cowan's suite of codes which can be found in the literature and by fully including the spin-orbit (S.O.) interaction in the 4f$^2$ ground electronic configuration and treating it as a perturbation in the 4f5d first-excited one.

The demonstration is made by re-examining the case of Pr$^{3+}$:YAG, a well known luminescent and laser material [13-25] with a very low energy 4f5d band [20, 22, 26]. It also gave the opportunity to provide more complete and more reliable absorption and emission data than provided in the past. This is of interest in order to re-examine the real laser potential of this material, Pr$^{3+}$:YAG being one of the rare Pr$^{3+}$ doped oxide crystals [27-37], compared with $Pr^{3+}$ doped fluorides, which gave rise to noticeable laser emissions in the green, orange and deep-red spectral domains [23-25].

The article is divided in three parts.

1) A first part aimed at gathering, improving and completing the already available spectroscopic data and answering a number of questions :

- that concerning the assignments of the absorption and emission lines based on the energy levels positions given in the literature,

- that concerning the purely radiative and fluorescence lifetimes of the $^3P_0$ and $^1D_2$ emitting levels, thus their quantum emission efficiencies at room temperature and for standard $Pr^{3+}$ ion concentrations (say between 0.5%at and 1%) for usual luminophor and laser applications,

- that concerning the coincidences between the main emission transitions originating from the principal $^3P_0$ emitting level with excited-state absorption transitions between this $^3P_0$ level ($4f^2$ fundamental configuration) and levels belonging to the 4f5d first-excited electronic configuration.

2) A second part devoted to the adequation of results obtained (i) from « standard » and « modified » (KM [8, 9]) Judd-Ofelt (J.O.) treatments between absorption spectra and emission data such as fluorescence lifetimes and branching ratios (not provided in the past literature), then (ii) from a new « extended » approach [11, 12] allowing in particular the possibility of emission transitions from the $^3P_0$ emitting level down to levels with odd J quantum numbers such as the $^3P_0 \rightarrow ^3H_5$ (around 560 nm) and $^3P_0 \rightarrow ^3F_3$ (around 730 nm). Such emission transitions are not allowed within the framework of standard J.O. theory while they can be characterized by experimentally measured branching ratios exceeding 10%, which is far from being negligible. A comparison is made with results obtained with $Pr^{3+}$:ZBLAN, a well-known fluorophosphate laser material which is characterized, as most of fluorides, by higher energy 4f5d energy levels, and whose luminescence properties were more specifically studied and detailed results can be found in the past literature [38, 39].

3) A third part addressing the question of the emission (and gain) cross sections of the above mentioned emission transitions which can be estimated from the registration of reliable emission spectra (with a good spectral resolution) and the estimation of reliable branching ratios, the latter being derived either experimentally (which assumes emission spectra corrected for the spectral response of the apparatuses in the whole spectral domain, which is not provided in the past literature) or being derived from the above « standard », « modified » and « extended » versions of J.O. theory.

## 2. Spectroscopic measurements

The optical absorption spectra were registered between 400 nm and 2500 nm by using a standard Perkin-Elmer (PE) double-beam Lambda 900 UV/VIS/NIR spectrophotometer.

Time-resolved emission spectra and fluorescence decays were registered by exciting the samples with an (EKSPLA NT342) OPO (Optical Parametric Oscillator) pumped by a frequency-tripled Nd:YAG pulsed laser delivering 10 Hz laser pulses of about 7 ns time duration and energies of several mJ. Measurements were then performed by exciting either into the $^3P_2$ excited level of $Pr^{3+}$ ions around 452 nm or into their $^1D_2$ level around 590 nm and by detecting and analyzing the emission signals originating from the $^3P_0$ and $^1D_2$ emitting levels in the visible spectral domain, between about 400 and 800 nm, and in the near-infrared up to about 1600 nm. For that purpose, use was made of a Shamrock 303i spectrometer with a 1200 grooves/mm grating followed by an i Star CCD or an InGaAs CCD ANDOR detector. All the emission spectra were corrected for the spectral response of the analysis and detection devices by using a calibrated light source.

## 2.1. Absorption spectra

The room temperature absorption spectrum of a 0.24%$Pr^{3+}$:YAG crystal (concentration determined par X-ray fluorescence analysis [17] which corresponds to about $3.29\times10^{19}$ $Pr^{3+}$ ions/cm$^3$) is shown in the Figure 1. It is similar to the spectrum reported by Cavalli et al [22], a spectrum, however, for which the thickness of the sample used is not indicated (which prevents the use of their absorption scale) and which suffers from a less reliable attribution of the lines in the near-infrared. The assignment reported in Fig. 1 is made by using the energy levels positions given in refs [13, 15, 18].

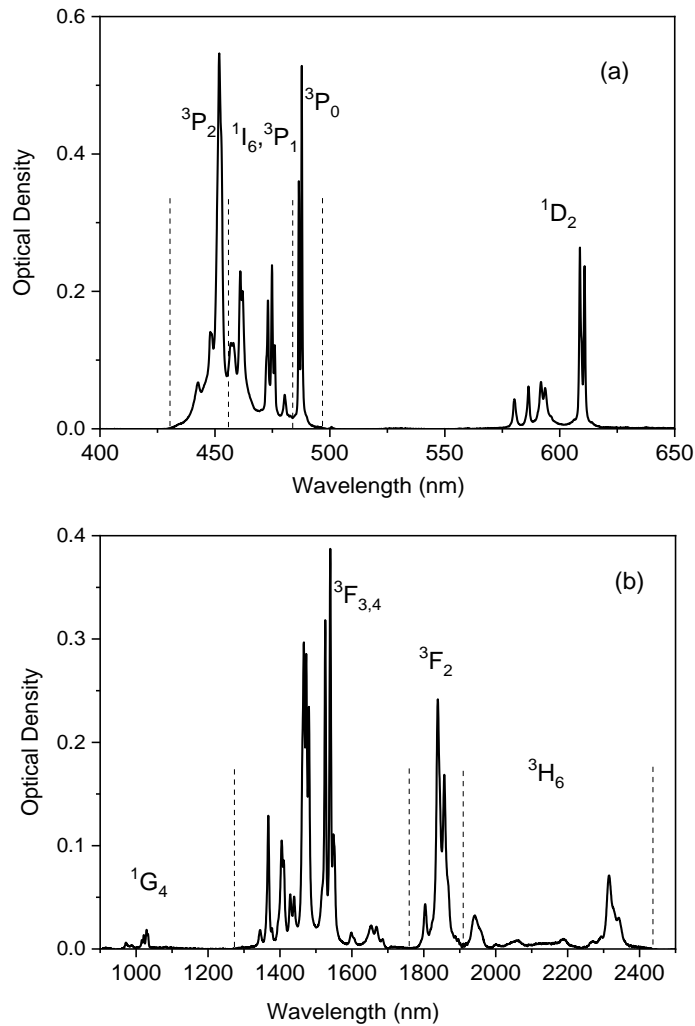

Figure 1 : Room temperature absorption (optical density) spectra of a 5.52 mm thick sample of 0.24%Pr :YAG crystal registered (a) in the visible and (b) the near- and mid-infrared spectral domains

## 2.2. Emission spectra

The time-resolved emission spectrum registered by exciting the sample at about 452 nm into the $^3H_4 \rightarrow ^3P_2$ absorption band (see in Fig. 1) and by collecting the emission signal with a short time delay (20 ns) and for a short time duration (10 µs) after the excitation laser pulse, is reported in the Figure 2. Because of these excitation and time-resolved emission conditions, the resulting spectrum can be safely attributed to emissions predominantly coming from the short-lived $^3P_0$ emitting level. In fact, what is indicated in Fig. 2, it must be attributed to emissions from both the $^3P_0$ emitting level and the above close-lying and thermalized $^3P_1$ and $^1I_6$ energy levels. The resulting spectrum notably differs from those reported in [22], i.e. spectra which present, by lack of time-resolution, a mix between emission lines originating from both the short-lived and long-lived $^3P_0$ and $^1D_2$ emitting levels. No clear distinction is made either in [22] with the lines coming from the $^3P_0$, $^3P_1$ and $^1I_6$ levels, and the band attributed to a combination of $^3P_1, ^3P_0, ^1I_6 \rightarrow ^1G_4$ emission transitions which appears between 850 nm and 1050 nm is not clearly deconvoluted.

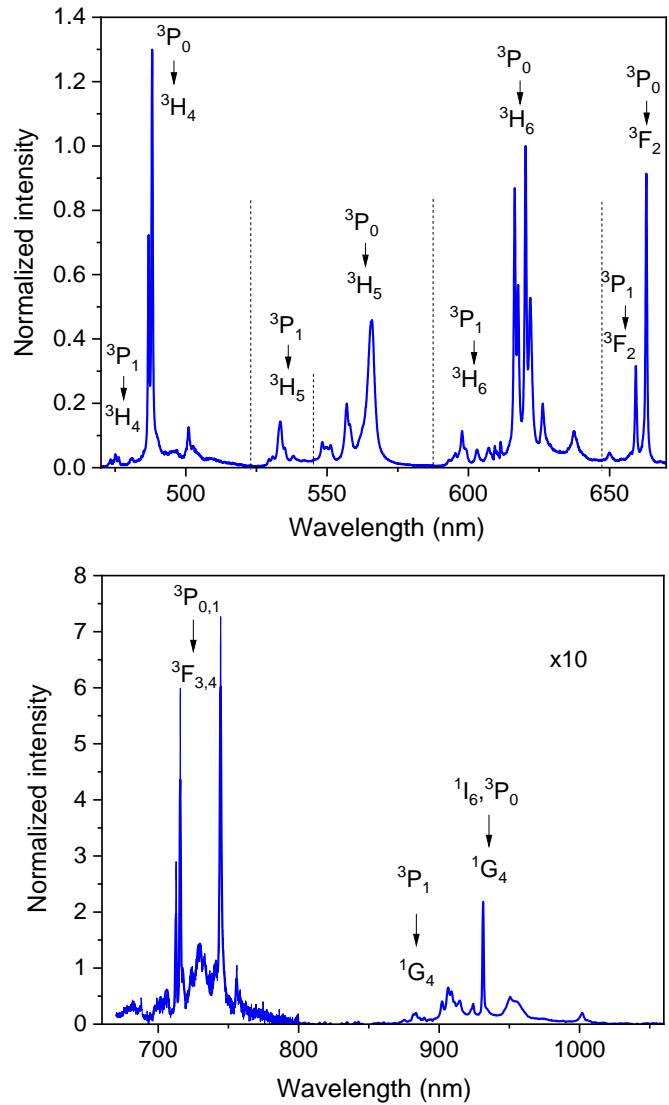

Figure 2 : Time-resolved room temperature emission spectra of $Pr^{3+}$:YAG from the $^3P_0$, $^3P_1$ and $^1I_6$ thermalized levels in the visible and near-infrared spectral range

The time-resolved emission spectrum registered by exciting the sample at about 590 nm into the $^3H_4 \rightarrow ^1D_2$ absorption band (see in Fig. 1) and by collecting the emission signal with a long time delay (60 µs) and for a long time duration (3 ms) after the excitation laser pulse is reported in the Figure 3. Doing so, the registered emission spectrum can be safely attributed to emissions from the long-lived $^1D_2$ emitting level essentially, which is not the case of the spectrum reported in [22]. Moreover, the obtained spectrum is much better resolved than that shown in [22].

It is also worth noting here that the above $^3P_0, ^3P_1, ^1I_6$ and $^1D_2$ emission spectra were carefully corrected from the spectral response of the apparatuses, which makes possible to compare the intensities of the various inter-mutiplet emission transitions and to derive reliable branching ratios. It is immediately seen from Fig.2, for instance, that the $^3P_0 \rightarrow ^3H_5$ and $^3P_0 \rightarrow ^3F_3$ inter-multiplet emission transitions occur with rather strong emission intensities, whereas they are assumed equal to zero within the « standard » J.O. formalism, a question which will be addressed more specifically in the theoretical analysis developed in Section 3.

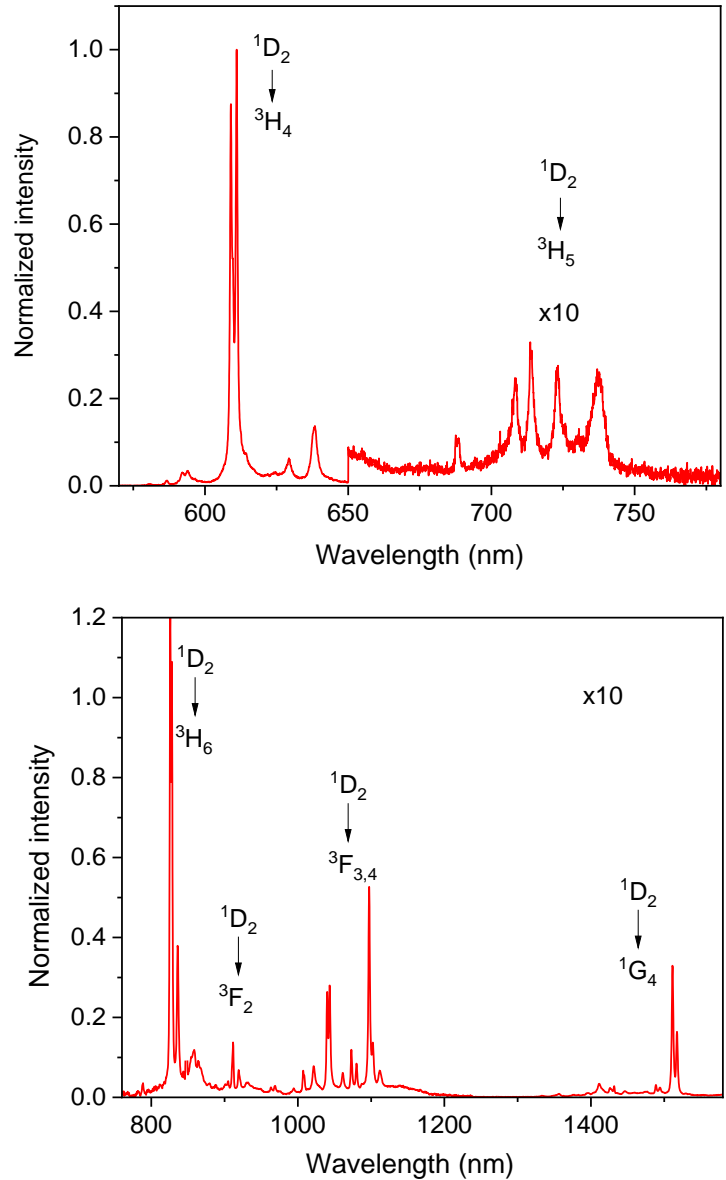

Figure 3 : Time-resolved room temperature emission spectra of Pr$^{3+}$:YAG from the $^1D_2$ emitting level in the visible and near-infrared spectral range

### 2.3 Fluorescence decays

Fluorescence decay measurements were performed at low (77K) and high (300K) temperatures by using a Pr$^{3+}$:YAG crystal with a very low dopant concentration of 0.08at%Pr$^{3+}$, and the results were compared with what was already reported in the literature, i.e. mainly in [17] for 0.08at%Pr$^{3+}$, in [16] for 0.12at%Pr$^{3+}$, in [24] for 0.6at%Pr$^{3+}$ and in [22] for 1at%Pr$^{3+}$ dopant concentrations. The $^3P_0$ fluorescence decays were registered by exciting the sample at 452 nm and monitoring the emission signal at 485 nm. They are reported in Figure 4(a). The decays are perfectly exponential with the decay time constants (fluorescence lifetimes) $\tau_F(^3P_0) \approx 11.9$ µs at 77K and 8.8 µs at 300K, in good agreement with what is reported in [16]. The $^1D_2$ fluorescence decays were registered by exciting the sample at 593 nm and monitoring the emission signal at 628 and 1516 nm (see in Fig. 3). They are displayed in Figure 4(b). The fluorescence decays are no longer perfectly but nearly exponential with the average decay time-constants $\tau_F(^1D_2) \approx 238$ µs at 77K and 196 µs at 300K, still in fairly good agreement with the results reported in [16].

The fact that all these fluorescence decays, especially those for the $^3P_0$ emitting level, are almost perfectly exponential simply indicates that, at a 0.08at%Pr$^{3+}$ dopant concentration, almost no inter-ionic energy transfers occur. The slight lengthening observed by lowering the temperature can be attributed either to decreasing non-radiative multiphonon relaxations or some change in the nature of the emitting levels. According to the analysis made in [40], it is demonstrated that the multiphonon relaxation rate of the excited energy levels of trivalent rare-earth ions in YAG is given by $W_{nr} = C.\exp(-\alpha.\Delta E)$ with C ≈ 6.7×10$^{13}$ s$^{-1}$ and α ≈ 7.5×10$^{-3}$ cm. It means that for the $^3P_0$ emitting level, with $\Delta E_{min}(^3P_0-^1D_2)$ ≈ 3302 cm$^{-1}$ [13, 15], the multiphonon rate is given by $W_{nr}$ ≈ 1176 s$^{-1}$ and $1/\tau_F(^3P_0)$ ≈ $1/\tau_R(^3P_0)$ + 1176, where $\tau_R(^3P_0)$ stands for the purely radiative lifetime of the $^3P_0$ level. Thus, since $\tau_F(^3P_0)$ ≈ 8.8 μs or 11.9 μs, it means that $\tau_F(^3P_0)$ ≈ $\tau_R(^3P_0)$ (within the experimental uncertainties). Therefore, the $^3P_0$ emission quantum efficiency is nearly equal to 1 and the effect of temperature needs to be assigned to a change in the nature of the emitting level. In fact, it is due to the contribution of the above-lying and thermalized $^3P_1$ et $^1I_6$ emitting levels, which will be included in the calculation of a so-called "effective" radiative lifetime and more specifically discussed in the following theoretical section. In the case of the $^1D_2$ emitting level, the multiphonon relaxation rate is obtained by using $\Delta E_{min}(^1D_0-^1G_4)$ ≈ 6064 cm$^{-1}$ which gives $W_{nr}$ ≈1771×10$^{-4}$ s$^{-1}$, thus a value which is nearly 10$^4$ smaller than the previous one. Therefore, in the case of $^1D_2$, even more than in the case of $^3P_0$, non-radiative multiphonon relaxations cannot be invoked to account for the observed temperature effect. It is assumed to be due to some residual inter-ionic and phonon-assisted energy transfers.

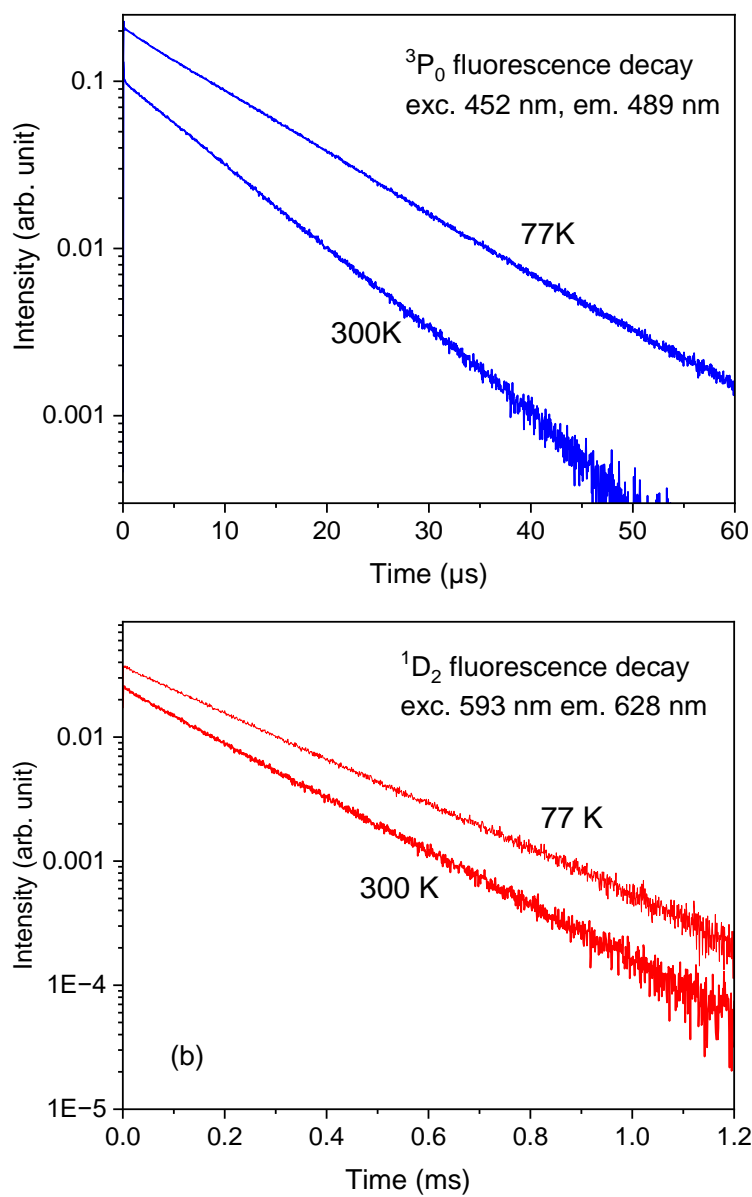

Figure 4 : Room (300 K) and low (77 K) temperature fluorescence decays (semi-log scale) of the $^3P_0$ and $^1D_2$ emitting levels of a 0.08at%Pr$^{3+}$:YAG crystal

## 2.4. Cross section spectra and branching ratios

The above registered and calibrated absorption and emission spectra can be transformed into cross section spectra using different techniques and can be used, along with fluorescence data, to derive inter-multiplet emission branching ratios.

For that purpose, let remember first, concerning the $^3P_0$ emitting level, the connection (Einstein reciprocity) which can be made between the $^3H_4 \to {^3P_0}$ absorption and $^3P_0 \to {^3H_4}$ emission spectra by using the well-known « Reciprocity » (or McCumber) and « Fuchtbauer-Ladenburg » methods. The former, which will be labelled « RP method », consists in transforming the $^3H_4 \to {^3P_0}$ absorption cross section spectrum into a stimulated emission one by using the expression [41]:

$$\sigma_{em}^{RP, u\to l}(\lambda) = \sigma_{abs}^{l\to u}(\lambda) \cdot \frac{Z_l}{Z_u} \cdot exp\left[\left(\frac{1}{\lambda_{ZL}} - \frac{1}{\lambda}\right) hc/kT\right] \quad (1)$$

in which $u$ and $l$ stand for the upper and lower levels of the considered emission transition, i.e. $u = {^3P_0}$ and $l = {^3H_4}$, where $\lambda_{ZL}$ (in nm unit) is the so-called "zero-line" (ZL) wavelength which corresponds to the energy difference between the lowest sub-levels of each multiplet of the transition, and $Z_m = \sum_k g_k^m \exp(-E_k^m/kT)$ is the partition function of the multiplet $m$ resulting from the energetic positions $E_k^m$ of the Stark levels of this multiplet and of their degeneracy $g_k^m$ (equal to 1 for $Pr^{3+}$:YAG). According to the energy level positions reported for $Pr^{3+}$:YAG in [13, 15] $Z_l = Z_{3H4} = 3.049$, $Z_u = Z_{3P0} = 1$ and $\lambda_{ZL} \approx 486.9$ nm.

Starting with the absorption spectrum reported in Fig. 1(a) and transforming it from optical density O.D. into absorption cross section $\sigma_{abs}$ using the expression $\sigma_{abs} = \frac{O.D.\ln 10}{N.d}$ where N stands for the $Pr^{3+}$ ion concentration, i.e. $3.29 \times 10^{+19}$ cm$^{-3}$ for 0.24at%$Pr^{3+}$, and d = 0.552 mm for the sample thickness, it is found, concerning the $^3H_4 \to {^3P_0}$ transition around 488 nm, the absorption cross section spectrum reported in Fig. 5(a). Then by using Expr. (1) this absorption cross section spectrum can be transformed into the $\sigma_{em}^{RP}$ emission cross section spectrum also reported in Fig. 5(a).

Now, the portion of emission spectrum reported in Fig. 2 between 480 nm and 520 nm and corresponding to the $^3P_0 \to {^3H_4}$ transition can be transformed into a $\sigma_{em}^{FL}$ emission cross section spectrum by using the « Fuchtbauer-Ladenburg » method labelled « FL » and given by the expression [41]:

$$\sigma_{em}^{FL, u\to l}(\lambda) = \frac{\lambda^5 \beta_R}{8\pi c n^2 \tau_R} \frac{I_{em}^{u\to l}(\lambda)}{\int \lambda I_{em}^{u\to l}(\lambda) d\lambda} \quad (2)$$

which applies for cubic as well as for isotropic crystals (no polarization effect), which is the case of $Pr^{3+}$:YAG. In this expression, $I_{em}$ stands for the registered emission intensity, $\beta_R$ for the branching ratio of the considered emission transition and $\tau_R$ for the radiative lifetime of the emitting level. According to fluorescence decay measurements (subsection 2.3), choice can be made for a radiative lifetime $\tau_R(^3P_0) = (10\pm1)$ μs. The branching ratio for the $^3P_0 \to {^3H_4}$ emission transition can be estimated from the emission spectra reported in Fig. 2 and the experimental values which can be calculated with the expression $\beta_{exp}^{u\to l} = \frac{\int \lambda . I_{u\to l}(\lambda) . d\lambda}{\sum_l \lambda . I_{u\to l}(\lambda) . d\lambda}$ (where $I_{i\to j}(\lambda)$ stand for the measured intensities and the integrations are made, in the case of the emitting level $^3P_0$, by including the contribution of the $^3P_1$ and $^1I_6$ emitting levels) thus $\beta_{exp}^{3P0 \to 3H4} \approx 13\%$. An alternative method to this FL method, also called FL $\beta/\tau$ method, consists in applying the same expression (2) but using the $\beta_R/\tau_R$ value which gives the best agreement with the emission spectrum derived with the RP method at the ZL wavelength. Using this $\beta/\tau$ method, the best agreement - see in Fig. 5 (a) - is obtained, keeping $\tau_R(^3P_0) = (10\pm1)$ μs, with $\beta_{exp}^{3P0 \to 3H4} \approx 26\%$, thus about twice the experimental value derived from the emission spectra. One reason for that is a well-known and unavailable reabsorption effect coming from the coincidences between absorption and emission lines and their high cross sections. This would also explain the different $\tau_F(^3P_0)$ fluorescence lifetimes, from 8.5 to 13 μs, reported in the past literature [17, 22-24] independently of the temperature (between 4 and 300K) and the Pr$^{3+}$ ion concentration (between 0.08% and 1%). Another reason could be due to an inadequacy of the $Z_{3H4}$ value used in the RP method in the case of Pr$^{3+}$:YAG. Indeed, as it was discussed in the past, for instance in [6], in systems like Pr$^{3+}$:YAG where multiplets are characterized by strong crystal field splittings, the effective ratio between partition functions or multiplet degeneracies which is used in expressions relating absorption (oscillator strength) and (stimulated) emission data can be smaller than theoretically expected. Therefore, at this step of our investigation, it is difficult to make a choice between the different methods and the question will be discussed further in the following theoretical analysis section.

We also report in Fig. 5 (b), which will be discussed more specifically in Section 4, the $^3P_0 \rightarrow\,^3H_4$ gain cross section spectra which can be derived by using the expression:

$$g(\lambda) = \beta^{exc}\sigma_{em}(\lambda) - (1-\beta^{exc})\sigma_{abs}(\lambda) \qquad (3)$$

where $\sigma_{abs}(\lambda)$ and $\sigma_{em}(\lambda)$ stand for the absorption and emission cross sections and $\beta^{ex} = N^{exc}/N_{tot}$ ($N_{tot}$ being the density of Pr$^{3+}$ ions and $N^{exc}$ the density of ions brought in the $^3P_0$ excited state) for the excitation ratio which can vary, depending on the excitation conditions, from 0 to 100%.

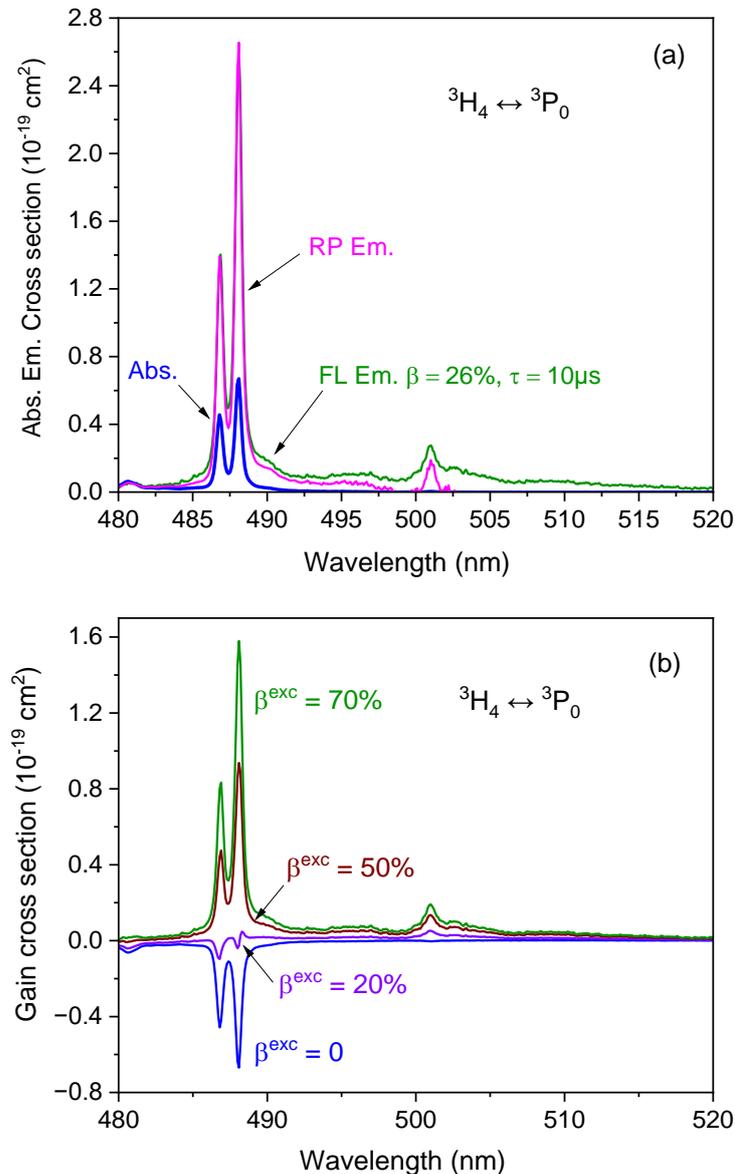

Figure 5 : (a) $^3H_4 \rightarrow\,^3P_0$ absorption and $^3P_0 \rightarrow\,^3H_4$ emission cross section spectra derived using Reciprocity (RP) and Fuchtbauer-Ladenburg (FL) methods ; (b) Resulting gain cross section spectra for various excitation ratios $\beta^{ex}$

### 3. Theoretical analysis

This part of the article is devoted to the analysis of the above spectroscopic data (absorption and emission intensities, emission lifetimes and branching ratios) within the framework of the « standard » J.O. theory and by using « modified » and « extended » versions of this formalism.

#### 3.1. Standard J.O. analysis

Based on the transition strengths of the various absorption transitions reported in Fig. 1, fit of the experimental data within the framework of « standard » J.O. theory (see for instance in [42] for a detailed description of the fitting procedure) led, by using the YAG dispersion relation which can be found in [43] and the $U^{(t)}$ electric-dipole transition matrix elements given by Weber [44] (except for the $^3H_4$-$^1G_4$ transition, for which use was made of the transition matrix elements given in our Cowan code, i.e. $U^{(2)}$ = 0.0018, $U^{(4)}$ = 0.0043 and $U^{(6)}$ = 0.0118), to the following results :

1. Without including in the fit the « hypersensitive » $^3H_4 \rightarrow ^3P_2$ absorption transition and by making the adjustment between the experimental and theoretical electric-dipole transition strengths usually noted $S_{ed}$ (see in [42] for detailed expressions), the best fit to the data is obtained with $\Omega_2$ = - 4.04, $\Omega_4$ = 10.08, $\Omega_6$ = 3.78 (in $10^{-20}$ cm$^2$), and with the RMS deviation $\delta$ = 0.714 (in $10^{-20}$ cm$^2$), thus with a negative (non-physical) $\Omega_2$ parameter, as it is often found with Pr$^{3+}$ doped materials and already reported by Malinowski et al [14] in the case of Pr$^{3+}$ :YAG.

2. Now including in the fit the « hypersensitive » $^3H_4 \rightarrow ^3P_2$ absorption transition and by making again the adjustment between the experimental and theoretical electric-dipole transition strengths $S_{ed}$, the best fit to the data is obtained with $\Omega_2$ = - 4.10, $\Omega_4$ = 9.95, $\Omega_6$ = 4.08 (en $10^{-20}$ cm$^2$), and with $\delta$ = 1.33 (en $10^{-20}$ cm$^2$), thus with nearly the same J.O. parameters as previously but with a less favorable RMS deviation.

So, let's examine in the following sections what is obtained within the framework of « modified » and « extended » versions of J.O. theory for which use is made, as we shall see, of positive parameters.

#### 3.2. Modified J.O. analysis

A modified version of the J.O. formalism, hereafter also labelled « **KM** » J.O. approach, was proposed in the past by Kornienko et al [8, 9] to better account for the influence of the 4f5d excited-state configuration on the electric-dipole transition strengths $S_{ed}$ by writing :

$$S_{ed}^{calc}(J \leftrightarrow J') = \sum_{t=2,4,6} \Omega_t \left| \langle \alpha J \| U^{(t)} \| \beta J' \rangle \right|^2 . \left[ 1 + 2\alpha(E_J + E_{J'} - 2E_f^0) \right] \quad (3)$$

where $E_J$ (taken equal to 0) stands for the energy of the $^3H_4$ fundamental level, $E_{J'}$ for the energy of a particular excited multiplet, $E_f^0$ for the energy corresponding to the center of gravity of the 4f$^2$ fundamental configuration (taken equal to about 9940 cm$^{-1}$), and $\alpha$ is an additional parameter which should be equal to 1/2$\Delta$fd where $\Delta$fd roughly corresponds to the energy difference between the levels of the 4f5d first excited configuration which give the largest contribution in the considered Pr$^{3+}$-doped material and the average energy $E_f^0$ [5].

Thus, including in the fit the « hypersensitive » $^3H_4 \rightarrow ^3P_2$ absorption transition and by making again the adjustment between the experimental and above defined theoretical electric-dipole transition strengths, the best fit to the data (see in Table 1) is obtained by taking $\alpha$ = 2.067×10$^{-5}$ cm, thus an energy difference $\Delta$fd ≈ E(4f5d) -E(4f) = 24189 cm$^{-1}$, and an excited configuration energy E(4f5d) ≈ 34129 cm$^{-1}$ which corresponds to the energy of the maximum of the lowest 4f5d band observed around 293 nm in Pr$^{3+}$ :YAG [20], and which can be also determined by using the relation and the data reported in the article by Dorenbos [26]. This fit is obtained with the J.O. parameters $\Omega_2$ = 1.644, $\Omega_4$ = 9.055 and $\Omega_6$ = 12.691 (in 10$^{-20}$ cm$^2$) and the RMS deviation $\delta$ = 0.77 (in 10$^{-20}$ cm$^2$), therefore with a positive $\Omega_2$ parameter and a better RMS than with the

« standard » approach. As shown in Table 1, the agreement between experimental and theoretical values is not very good, especially for the transitions to $^3P_0$ and $^1D_2$, but also for the transitions to $^3H_6$ and $^3P_2$.

| Abs transition | Wavelength (nm) | $S_{edexp}$ ($10^{-20}$ cm$^2$) | $S_{edcalc}$ ($10^{-20}$ cm$^2$) | Ratio (cal/exp) |
|---|---|---|---|---|
| $^3H_4 \rightarrow ^3H_6$ | 2200.00 | 1.503 | 0.7605 | 0.506 |
| $^3H_4 \rightarrow ^3F_2$ | 1839.00 | 2.4856 | 2.4244 | 0.975 |
| $^3H_4 \rightarrow ^3F_{3,4}$ | 1494.00 | 8.088 | 8.5861 | 1.061 |
| $^3H_4 \rightarrow ^1G_4$ | 1000.00 | 0.274 | 0.1806 | 0.659 |
| $^3H_4 \rightarrow ^1D_2$ | 600.00 | 1.2857 | 0.6731 | 0.523 |
| $^3H_4 \rightarrow ^3P_0$ | 488.00 | 0.9282 | 1.5906 | 1.713 |
| $^3H_4 \rightarrow ^3P_1+^1I_6$ | 473.00 | 2.9205 | 2.351 | 0.804 |
| $^3H_4 \rightarrow ^3P_2$ | 452.00 | 3.5207 | 2.2604 | 0.642 |

Table 1. Absorption transitions, average transition wavelengths, experimental and calculated electric-dipole transition strengths according to the « modified » (KM) J.O. approach

The above obtained J.O. parameters lead in turn to the radiative lifetimes $\tau_R(^3P_0) \approx 9.9$ μs, $\tau_R(^1D_2) \approx 175$ μs and $\tau_R(^1G_4) \approx 1.16$ ms, the first two values being rather close to the above reported fluorescence lifetimes $\tau_F(^3P_0) \approx 8.8$ μs or 11.9 μs and $\tau_F(^1D_2) \approx 196$ μs or 238 μs, depending on the temperature.

From this point of view, the $^3P_0$ emission lifetime measured at room temperature should be compared in fact to some effective radiative lifetime resulting from the thermalization and the contribution of the above close-lying $^3P_1$ et $^1I_6$ emitting levels. Such effective lifetime is given by:

$$\tau^{eff}_{3P0} = \frac{g_{3P0}+(g_{3P1}+g_{1I6})\,e^{-\Delta E/kT}}{g_{3P0}\sum_{J'} A^{em}_{3P0-J'}+(g_{3P1}\sum_{J'} A^{em}_{3P1-J'}+g_{1I6}\sum_{J'} A^{em}_{1I6-J'})e^{-\Delta E/kT}} \quad (4)$$

an expression which was often used in the past (see for instance in [39] and [45]) by assuming that the $^3P_0$ level is thermalized with $^3P_1$ et $^1I_6$ levels located above at an average energy ΔE corresponding to the lowest Stark component of the $^3P_1$ multiplet, i.e. ΔE = 522 cm$^{-1}$ in the case of Pr$^{3+}$:YAG [13, 15]. $g_{3P0} = 1$, $g_{3P1} = 3$ and $g_{1I6} = 13$ stand for the multiplet degeneracies and the summation terms appearing in the denominator stand for the inverse of the radiative lifetimes of the various multiplets derived in the J.O. treatment. For instance, by using the data of the above « modified » J.O. analysis, i.e. $\sum_{J'} A^{em}_{3P0-J'} = 1/\tau_R(^3P_0)$, $\sum_{J'} A^{em}_{3P1-J'} = 1/\tau_R(^3P_1)$, and $\sum_{J'} A^{em}_{1I6-J'} = 1/\tau_R(^1I_6)$, with $\tau_R(^3P_1) = 9.45$ μs and $\tau_R(^1I_6) = 36.7$ μs, it is found at room temperature, i.e. for kT ≈ 209 cm$^{-1}$, a $^3P_0$ effective radiative lifetime $\tau^{eff}_{3P0}$ = 14.8 μs, thus a bit too long compared with the measured fluorescence lifetime of about 10 μs.

Table 2 and Table 3 give the results obtained for the branching ratios of the $^3P_0$ and $^1D_2$ emission transitions, i.e. :
- the nature of the considered inter-multiplet emission transitions
- their average emission transition wavelengths
- the calculated (electric-dipole + magnetic-dipole) emission transition rates noted A$_{em}$
- the sum of these transition rates noted A$_{emtot}$, the ones give the calculated radiative lifetimes $\tau_R$ = (A$_{emtot}$)$^{-1}$
- the calculated radiative branching ratios β$_{calc}$ = A$_{em}$/A$_{emtot}$
- the experimental branching ratios determined from the spectra given in Fig. 2 and Fig. 3 and by assuming in the case of the emitting level $^3P_0$ (see in section 2.4) a value of 26% for the $^3P_0 \rightarrow ^3H_4$ emission transition

- the RMS standard deviation between experimental and calculated branching ratios noted $\Delta\beta_i = \sqrt{\sum_j \frac{(\beta_{exp}^{eff}(i \to j) - \beta_{calc}(i \to j))^2}{p}}$, where p is the number of considered inter-multiplet transitions and i stands for $^3P_0$ and $^1D_2$

and, in the case of the $^3P_0$ emission transitions:

- the effective radiative branching ratios $\beta^{eff}$ which can be calculated, as noticed in Section 2.3, by including the contribution of the thermalized $^3P_1$ and $^1I_6$ emitting levels and by using an expression similar to that used to derive the above $\tau_{3P0}^{eff}$ value, i.e.:

$$\beta_{3P0-J}^{eff} = \frac{g_{3P0} A_{3P0-J}^{em} + (g_{3P1} A_{3P1-J}^{em} + g_{1I6} A_{1I6-J}^{em}) e^{-\Delta E/kT}}{g_{3P0} \sum_{J'} A_{3P0-J'}^{em} + g_{3P1} \sum_{J'} A_{3P1-J'}^{em} e^{-\Delta E1/kT} + g_{1I6} \sum_{J'} A_{1I6-J'}^{em} e^{-\Delta E2/kT}} \quad (5)$$

and, at the end:

- the corresponding RMS standard deviation between calculated and experimentally measured branching ratios by using the expression $\Delta\beta_i^{eff} = \sqrt{\sum_j \frac{\left[\beta_{exp}^{eff}(i \to j) - \beta_{calc}^{eff}(i \to j)\right]^2}{p}}$

| $^3P_0$ Emission | Wavelength (nm) | $A_{em}$ (s$^{-1}$) | $A_{emtot}$ (s$^{-1}$) | $\beta_{calc}$ Branching ratio (%) | $\beta_{exp}^{eff}$ Branching ratio (%) | $\Delta\beta_{3P0}$ (%) | $\beta_{calc}^{eff}$ Branching ratio (%) | $\Delta\beta_{3P0}^{eff}$ (%) |
|---|---|---|---|---|---|---|---|---|
| $^3P_0 \to {}^3H_4$ | 490 | 5.82 10$^4$ | 1.01 10$^5$ | 57.6 | 26 | 15,1 | 42.2 | 8.0 |
| $^3P_0 \to {}^3H_5$ | 560 | 0 | | 0 | 13.8 | | 6.3 | |
| $^3P_0 \to {}^3H_6$ | 618 | 1.97 10$^4$ | | 19.5 | 21.8 | | 14.9 | |
| $^3P_0 \to {}^3F_2$ | 662 | 8.59 10$^3$ | | 8.51 | 5.6 | | 6.43 | |
| $^3P_0 \to {}^3F_3$ | 715 | 0 | | 0 | 22.4 | | 2.81 | |
| $^3P_0 \to {}^3F_4$ | 744 | 1.16 10$^4$ | | 11.5 | | | 15.0 | |
| $^3P_0 \to {}^1G_4$ | 931 | 2.77 10$^3$ | | 2.75 | 10.4 | | 10.1 | |
| $^3P_0 \to {}^1D_2$ | 2576 | 8.54 | | 0.008 | ? | | 0.5 | |

Table 2. $^3P_0 \to {}^3H_4, {}^3H_5, {}^3H_6, {}^3F_2, {}^3F_3, {}^3F_4, {}^1G_4, {}^1D_2$ emission transition rates, branching ratios and RMS standard deviations between calculated and experimental branching ratios, including and not including the contribution of the $^3P_1$ and $^1I_6$ emitting levels, according to the « modified » (KM) J.O. approach

| $^1D_2$ Emission | Wavelength (nm) | $A_{em}$ (s$^{-1}$) | $A_{emtot}$ (s$^{-1}$) | $\beta_{calc}$ Branching ratio % | $\beta_{exp}$ Branching ratio exp % | $\Delta\beta_{1D2}^{eff}$ (%) |
|---|---|---|---|---|---|---|
| $^1D_2 \to {}^3H_4$ | 611 | 2460.1 | 5705.01 | 43.1 | 57 | 10.2 |
| $^1D_2 \to {}^3H_5$ | 722 | 44.99 | | 0.79 | 7.9 | |
| $^1D_2 \to {}^3H_6$ | 841 | 946.48 | | 16.59 | 14.3 | |
| $^1D_2 \to {}^3F_2$ | 912 | 877 | | 15.37 | 3.2 | |
| $^1D_2 \to {}^3F_3$ | 1039 | 167.8 | | 2.94 | | |
| $^1D_2 \to {}^3F_4$ | 1096 | 566.45 | | 9.92 | 12.4 | |
| $^1D_2 \to {}^1G_4$ | 1500 | 642.19 | | 11.79 | 5 | |

Table 3. $^1D_2 \to {}^3H_4, {}^3H_5, {}^3H_6, {}^3F_2, {}^3F_3, {}^3F_4, {}^1G_4$ emission transition rates, branching ratios and RMS standard deviations between calculated and experimental branching ratios, according to the « modified » (KM) J.O. approach

According to the results reported in Table 2 for the $^3P_0$ emitting level, it is first noticed, what is inherent to standard J.O. theory, that no emission transition rate is predicted for transitions $^3P_0 \rightarrow {}^3H_5$ and $^3P_0 \rightarrow {}^3F_3$ whereas it is found experimentally a branching ratio of at least 13.8 % for the former and, probably, according to Fig. 2, around 10 % for the latter. Moreover, except for transitions $^3P_0 \rightarrow {}^3H_6$ and $^3P_0 \rightarrow {}^3F_2$, the agreement between experimental and calculated branching for transitions $^3P_0 \rightarrow {}^3H_4$, $^3P_0 \rightarrow {}^3F_4$ and $^3P_0 \rightarrow {}^1G_4$ is rather poor, with a clear improvement, when considering the thermalization of the $^3P_0$, $^3P_1$ and $^1I_6$ levels and by assuming a branching ratio of 26 % for the $^3P_0 \rightarrow {}^3H_4$ emission transition.

In the case of the $^1D_2$ emitting level, the agreement between calculated and experimental values remains rather poor, although more satisfactory than in the case of $^3P_0$, especially for the emission transitions down to levels $^3H_4$, $^3H_6$ and $^3F_{3,4}$.

### 3.3. Extended J.O. analysis

The current extension of the J.O. model is based on our previous works, published by Hovhannesyan et al [11, 12]. The electric-dipole transition strengths are calculated using perturbation theory including the corrections due to the crystal field on the free-ion eigenvectors of the initial and final multiplets of the transitions. The main novelty of the current work is that we go beyond the approximation according to which all the levels of the excited configuration 4f5d are degenerate. To account for the width of that configuration, we take in the sum-over-state formula (see in Ref. [12], first two lines of Eq. (4)) the energies of the free-ion levels shifted by a chosen quantity. This represents a rather simple and versatile way to account for the strong perturbation due to the surrounding crystal, especially YAG, on the free-ion levels of the excited configuration. This extension, denoted "EXT 2" is discussed in subsection 3.3.2. For comparison, we still analyze the situation where all the levels of the 4f5d configuration are assumed degenerate (method "EXT 1", subsection 3.3.1), but unlike our previous works, we take their energy as a tunable parameter.

Compared with Refs. [11,12], our current extensions account for overlapping transitions, i.e. with several inseparable upper multiplets like $^3P_1$ and $^1I_6$. Moreover, for all transitions, the magnetic-dipole contribution is calculated, and subtracted to the experimental oscillator (or transition) strengths, in order to obtain the electric-dipole experimental line strengths. The latter are compared to the theoretical ones given by:

$$S_{ed}^{theo}(J \leftrightarrow J') = a_0^2 \sum_{k=1,3,5} C_k(J \leftrightarrow J') X_k \qquad (6)$$

where $a_0$ = 5.2917×10$^{-9}$ cm stands for the Bohr radius, $X_k$ (with k = 1, 3, 5) are three adjustable parameters that only depend on the crystal field potential, and $C_k$ (J $\leftrightarrow$ J') are three series of odd electric-dipole transition matrix elements calculated from free-ion eigenvectors and accounting for the spin-orbit interaction responsible for spin-changing transitions. The free-ion eigenvectors were calculated using the Racah-Slater method present in the Cowan code [46], based on the experimental free-ion energy levels of the NIST ASD database [47]. The one-electron wave functions $P_{n\ell}(r)$ calculated by Cowan serve to calculate the transition integrals $\langle 4f|r^k|5d\rangle = \int dr P_{4f} r^k P_{5d}$, which in the case of Pr$^{3+}$ are respectively equal to 0.7853 $a_0$, 3.918 $a_0^3$ and 28.42 $a_0^5$ for k = 1, 3 and 5. The free-ion spectrum coming from Cowan serve as input in our home-made code "jo_so" (https://gitlab.com/labicb/joso), to adjust $S_{ed}^{theo}(J \leftrightarrow J')$ line strengths with experimental ones.

3.3.1. Extended J.O. approach (EXT 1)

Within this first approach which will be labelled « **EXT. 1** », the 4f5d first excited electronic configuration is assumed to occur with a single energy corresponding to the maximum of the lowest 4f5d absorption band, i.e. around 34 000 cm$^{-1}$ in the case of Pr$^{3+}$:YAG (just like in the case of the above modified KM approach). In fact, this value also corresponds, what is found by varying it from about 29 000 cm$^{-1}$ to 60 000 cm$^{-1}$ (see in Figure 5 of Section 4), to the energy which leads to the best fit to the experimental data. This fit is obtained for $X_1$ = 1.022×10$^{-5}$, $X_3$ = 1.067×10$^{-6}$ and $X_5$ = 2.656×10$^{-8}$ with a RMS deviation δ = 0.477×10$^{-20}$cm$^2$, thus better than previously.

The results (experimental and calculated electric-dipole oscillator strengths, and ratio between them) are reported in Table 4, knowing that electric-dipole oscillator and transition strengths are related by the expression:

$$f_{ed}(J \to J') = 1.0847 \cdot 10^{18} \frac{\chi_{ed}}{(2J+1)\lambda} S_{ed}(J \to J') \qquad (7)$$

with $\chi_{ed} = \frac{(n^2+2)^2}{9n}$ and n the refractive index.

| Abs transition | Wavelength (nm) | $f_{edexp}$ (10⁻⁶) | $f_{edcalc}$ (10⁻⁶) | Ratio (cal/exp) |
|---|---|---|---|---|
| $^3H_4 \to {}^3H_6$ | 2200.00 | 1.39 | 0.7844 | 0.56 |
| $^3H_4 \to {}^3F_2$ | 1839.00 | 2.77 | 2.753 | 0.99 |
| $^3H_4 \to {}^3F_{3,4}$ | 1494.00 | 11.1 | 11.426 | 1.03 |
| $^3H_4 \to {}^1G_4$ | 1000.00 | 0.57 | 0.330 | 0.58 |
| $^3H_4 \to {}^1D_2$ | 600.00 | 4.49 | 2.227 | 0.5 |
| $^3H_4 \to {}^3P_0$ | 488.00 | 4.03 | 5.581 | 1,39 |
| $^3H_4 \to {}^3P_1+{}^1I_6$ | 473.00 | 13.1 | 11.790 | 0.9 |
| $^3H_4 \to {}^3P_2$ | 452.00 | 16.6 | 15.992 | 0.97 |

Table 4. Absorption transitions, average transition wavelengths, experimental and calculated electric-dipole oscillator strengths, according to extended J.O. theory (EXT. 1 approach)

Compared with the previous « modified » KM method, the agreement between experimental and calculated values is a little less good for transition $^3H_4 \to {}^1G_4$ but it is much better for transitions $^3H_4 \to {}^3P_0$ and $^3H_4 \to {}^3P_2$. The above obtained $X_k$ parameters lead in turn to the radiative lifetimes $\tau_R(^3P_0) \approx 9.26$ µs, $\tau_R(^1D_2) \approx 173$ µs and $\tau_R(^1G_4) \approx 1.17$ ms, thus about the same values as that found with the KM approach. They also give the values $\tau_R(^3P_1) = 8.34$ µs, $\tau_R(^1I_6) = 30.5$ µs and, according to Expr. 2, the $^3P_0$ effective radiative lifetime $\tau_{3P0}^{eff} = 13.4$ µs, thus a slightly better value than that obtained with the KM approach.

As above, we also report in the tables 5 and 6, the results obtained for the transition rates and branching ratios of the various inter-multiplet emission transitions coming from the $^3P_0$ (thermalized or not with the $^3P_0$ and $^1I_6$) and $^1D_2$ levels.

| $^3P_0$ Emission | Wavelength (nm) | $A_{em}$ (s⁻¹) | $A_{emtot}$ (s⁻¹) | $\beta_{calc}$ Branching ratio % | $\beta_{exp}^{eff}$ Branching ratio % | $\Delta\beta_{3P0}$ (%) | $\beta_{calc}^{eff}$ Branching ratio % | $\Delta\beta_{3P0}^{eff}$ (%) |
|---|---|---|---|---|---|---|---|---|
| $^3P_0 \to {}^3H_4$ | 490 | 4.75 10⁴ | 1.08 10⁵ | 44 | 26 | 11.5 | 32.4 | 5.4 |
| $^3P_0 \to {}^3H_5$ | 560 | 4.60 10³ | | 4.26 | 13.8 | | 8.1 | |
| $^3P_0 \to {}^3H_6$ | 618 | 2.27 10⁴ | | 21 | 21.8 | | 15.8 | |
| $^3P_0 \to {}^3F_2$ | 662 | 1.99 10⁴ | | 18.4 | 5.6 | | 13.1 | |
| $^3P_0 \to {}^3F_3$ | 715 | 1.17 10³ | | 1.1 | 22.4 | | 3.7 | |
| $^3P_0 \to {}^3F_4$ | 744 | 9.81 10³ | | 9.1 | | | 16.0 | |
| $^3P_0 \to {}^1G_4$ | 931 | 2.38 10³ | | 2.2 | 10.4 | | 9.6 | |
| $^3P_0 \to {}^1D_2$ | 2576 | 21.7 | | 0.02 | ? | | 1 | |

Table 5. $^3P_0 \rightarrow {}^3H_4, {}^3H_5, {}^3H_6, {}^3F_2, {}^3F_3, {}^3F_4, {}^1G_4, {}^1D_2$ emission transition rates, branching ratios and RMS standard deviations between calculated and experimental branching ratios, including ($\Delta\beta_{3P0}^{eff}$) and not including ($\Delta\beta_{3P0}$) the contribution of the $^3P_1$ and $^1I_6$ emitting levels, according to extended J.O. theory (EXT. 1 approach)

The agreement between experimental and calculated branching ratios is better than with the "modified" Kornienko method (5.4% against 8.0%). This mainly comes from the better agreement of the branching ratio of the $^3P_0 \rightarrow {}^3H_4$ transition, which is the strongest one. Moreover, unlike the Kornienko method, the "EXT 1" one predicts non-zero branching ratios for the $^3P_0 \rightarrow {}^3H_5$ and $^3P_0 \rightarrow {}^3F_3$ transitions, although they remain smaller than the experimental values.

| $^1D_2$ Emission | Wavelength (nm) | $A_{em}$ (s$^{-1}$) | $A_{emtot}$ (s$^{-1}$) | $\beta_{calc}$ Branching ratio % | $\beta_{exp}$ Branching ratio % | $\Delta\beta_{1D2}$ (%) |
|---|---|---|---|---|---|---|
| $^1D_2 \rightarrow {}^3H_4$ | 611 | 2370 | 5780 | 41 | 57 | 10.0 |
| $^1D_2 \rightarrow {}^3H_5$ | 722 | 55.9 | | 0.96 | 7.9 | |
| $^1D_2 \rightarrow {}^3H_6$ | 841 | 596 | | 10.3 | 14.3 | |
| $^1D_2 \rightarrow {}^3F_2$ | 912 | 532 | | 9.2 | 3.2 | |
| $^1D_2 \rightarrow {}^3F_3$ | 1039 | 152 | | 2.62 | | |
| $^1D_2 \rightarrow {}^3F_4$ | 1096 | 1390 | | 24.1 | 12.4 | |
| $^1D_2 \rightarrow {}^1G_4$ | 1500 | 685 | | 11.8 | 5 | |

Table 6. $^1D_2 \rightarrow {}^3H_4, {}^3H_5, {}^3H_6, {}^3F_2, {}^3F_3, {}^3F_4, {}^1G_4$ emission transition rates, branching ratios and RMS standard deviations between calculated and experimental branching ratios, according to extended J.O. theory (EXT. 1 approach)

In the case of the $^1D_2$ emitting level (Table 6), the results are similar to that found with the « modified » KM method with about the same $A_{emtot}$ and $\Delta\beta_{1D2}$.

3.3.2. Extended J.O. approach (EXT 2)

This second extended J.O. approach which will be labelled « **EXT. 2** » consists in assuming the interaction with a full free-ion 4f5d first excited electronic configuration shifted to lower energies in order to have the lowest levels around the position of the lowest 4f5d absorption band. In that case, the best fit to the experimental data is obtained with an energy of about 32 000 cm$^{-1}$ (see in figure 5 of Section 4), the parameters $X_1$ = 1.616×10$^{-5}$, $X_3$ = 7.758×10$^{-7}$ and $X_5$ = 2.972×10$^{-8}$ and with a RMS deviation $\delta$ = 0.444×10$^{-20}$ cm$^2$. The results (experimental and calculated electric-dipole oscillator strengths and ratio between them) are reported in Table 7. The agreement remains poor for the transitions to the $^3H_6$, $^1G_4$ and $^1D_2$ levels but it is excellent for the transitions to the other levels. The obtained $X_k$ parameters lead to the radiative lifetimes $\tau_R(^3P_0) \approx 12.3$ µs, $\tau_R(^1D_2) \approx 178$ µs and $\tau_R(^1G_4) \approx 1.52$ ms. They also give the values $\tau_R(^3P_1) = 12.6$ µs, $\tau_R(^1I_6) = 42.0$ µs and $\tau_{3P0}^{eff} = 18.5$ µs, thus slightly larger values, but still reasonable compared with measured fluorescence lifetimes, than that found with the other approaches.

| Abs transition | Wavelength (nm) | $f_{edexp}$ (10$^{-6}$) | $f_{edcalc}$ (10$^{-6}$) | Ratio (cal/exp) |
|---|---|---|---|---|
| $^3H_4 \rightarrow {}^3H_6$ | 2200.00 | 1.39 | 0.740 | 0.53 |
| $^3H_4 \rightarrow {}^3F_2$ | 1839.00 | 2.77 | 2.74 | 0.99 |
| $^3H_4 \rightarrow {}^3F_{3,4}$ | 1494.00 | 11.1 | 11.5 | 1.05 |
| $^3H_4 \rightarrow {}^1G_4$ | 1000.00 | 0.57 | 0.228 | 0.40 |
| $^3H_4 \rightarrow {}^1D_2$ | 600.00 | 4.49 | 2.57 | 0.57 |
| $^3H_4 \rightarrow {}^3P_0$ | 488.00 | 4,03 | 4.00 | 0.99 |
| $^3H_4 \rightarrow {}^3P_1+{}^1I_6$ | 473.00 | 13.1 | 12.96 | 0,99 |
| $^3H_4 \rightarrow {}^3P_2$ | 452.00 | 16.6 | 15.1 | 0,91 |

Table 7. Absorption transitions, average transition wavelengths, experimental and calculated electric-dipole oscillator strengths, according to extended J.O. theory (EXT. 2 approach)

The comparison between experimental and calculated branching ratios is reported in Table 8 for $^3P_0$ and Table 9 for $^1D_2$.

| $^3P_0$ Emission | Wavelength (nm) | $A_{em}$ (s$^{-1}$) | $A_{emtot}$ (s$^{-1}$) | $\beta_{calc}$ Branching ratio % | $\beta_{exp}^{eff}$ Branching ratio % | $\Delta\beta_{3P0}$ (%) | $\beta_{calc}^{eff}$ Branching ratio % | $\Delta\beta_{3P0}^{eff}$ (%) |
|---|---|---|---|---|---|---|---|---|
| $^3P_0 \to {}^3H_4$ | 490 | 3.40 10$^4$ | 7.97 10$^4$ | 42.9 | 26 | 13.5 | 41.8 | 10.2 |
| $^3P_0 \to {}^3H_5$ | 560 | 3.77 10$^3$ | | 4.75 | 13.8 | | 9.22 | |
| $^3P_0 \to {}^3H_6$ | 618 | 1.46 10$^4$ | | 18.5 | 21.8 | | 16.8 | |
| $^3P_0 \to {}^3F_2$ | 662 | 2.05 10$^4$ | | 25.8 | 5.6 | | 23.5 | |
| $^3P_0 \to {}^3F_3$ | 715 | 9.71 10$^2$ | | 1.22 | 22.4 | | 4.45 | |
| $^3P_0 \to {}^3F_4$ | 744 | 5.06 10$^3$ | | 6.38 | | | 16.4 | |
| $^3P_0 \to {}^1G_4$ | 931 | 6.80 10$^2$ | | 0.86 | 10.4 | | 7.26 | |
| $^3P_0 \to {}^1D_2$ | 2576 | 25.5 | | 0.03 | ? | | 0.86 | |

Table 8. $^3P_0 \to {}^3H_4$, $^3H_5$, $^3H_6$, $^3F_2$, $^3F_3$, $^3F_4$, $^1G_4$, $^1D_2$ emission transition rates, branching ratios and RMS standard deviations between calculated and experimental branching ratios, including and not including the contribution of the $^3P_1$ and $^1I_6$ emitting levels, according to extended J.O. theory (EXT 2 approach)

| $^1D_2$ Emission | Wavelength (nm) | $A_{em}$ (s$^{-1}$) | $A_{emtot}$ (s$^{-1}$) | $\beta_{calc}$ Branching ratio % | $\beta_{exp}$ Branching ratio % | $\Delta\beta_{1D2}$ (%) |
|---|---|---|---|---|---|---|
| $^1D_2 \to {}^3H_4$ | 611 | 2740 | 5620 | 53.6 | 57 | 9.3 |
| $^1D_2 \to {}^3H_5$ | 722 | 38.3 | | 0.75 | 7.9 | |
| $^1D_2 \to {}^3H_6$ | 841 | 282 | | 5.52 | 14.3 | |
| $^1D_2 \to {}^3F_2$ | 912 | 565 | | 11.1 | 3.2 | |
| $^1D_2 \to {}^3F_3$ | 1039 | 143 | | 2.79 | | |
| $^1D_2 \to {}^3F_4$ | 1096 | 1380 | | 27.1 | 12.4 | |
| $^1D_2 \to {}^1G_4$ | 1500 | 469 | | 9.19 | 5 | |

Table 9. $^1D_2 \to {}^3H_4$, $^3H_5$, $^3H_6$, $^3F_2$, $^3F_3$, $^3F_4$, $^1G_4$ emission transition rates, branching ratios and RMS standard deviations between calculated and experimental branching ratios, according to extended J.O. theory (EXT 2 approach)

Regarding the $^3P_0$ emitting level, the "EXT 2" method gives a poorer agreement than "EXT 1", which is visible on the strongest $^3P_0 \to {}^3H_4$ emission transition, but also on other ones like $^3P_0 \to {}^3F_2$. Regarding $^1D_2$, the agreement for the strongest transition to $^3H_4$ is better than for "KM" and "EXT 1", but it is poor with the three methods for the other transitions.

### 3.4. Comparison of theoretical approaches

The adjusting parameters found with the three different J.O. approaches (KM, EXT1 and EXT2) are reported in Table 10 for sake of clarity. The $\Omega_{2,4,6}$ parameters are given in 10$^{-20}$ cm$^2$. The $X_{1,3,5}$ are given in atomic units, i.e. $(E_h/a_0^k)^2$, where $E_h$ is the Hartree energy and $a_0$ the Bohr radius. The RMS standard deviation $\delta$ (in 10$^{-20}$ cm$^2$) is calculated from the electric dipole transition strengths (by subtracting the calculated magnetic-dipole contributions). The relative standard deviation $\delta_{rel}$ is calculated by dividing the $\delta$ value by the largest measured electric-dipole transition strength.

As reported above (See in Section 3.2) the best agreement between experimental and calculated oscillator (or transition) strengths in the case of the KM « modified » J.O. approach is obtained for a first-excited configuration energy E(4f5d) = 34 000 cm$^{-1}$ [20, 26]. In fact, this energy represents the energy for which it is obtained a good agreement between experimental and calculated values together with the most significant $^3P_0$ and $^1D_2$ radiative lifetimes. Indeed, with this KM method, there is no minimum E(4f5d) value for which it is obtained both a minimum RMS value and significant radiative lifetimes : the RMS continues to decrease as the value of E(4f5d) is decreased, even for values that no longer have physical significance, and the calculated lifetimes fall well below those measured experimentally.

| Parameters (10$^{-20}$ cm$^2$) | JO modified (KM) approach (with E4f5d = 34000 cm$^{-1}$) | Parameters (a.u.) | JO Extended theory (EXT. 1) (flat with E4f5d = 34 000 cm$^{-1}$) | JO Extended theory (EXT. 2) (full 4f5d shifted down to 32 000 cm$^{-1}$) |
|---|---|---|---|---|
| $\Omega_2$ | 1.644 | $X_1$ | 1.022×10$^{-5}$ | 1.616×10$^{-5}$ |
| $\Omega_2$ | 9.055 | $X_3$ | 1.067×10$^{-6}$ | 7.758×10$^{-7}$ |
| $\Omega_2$ | 12.691 | $X_5$ | 2.656×10$^{-8}$ | 2.972×10$^{-8}$ |
| δ (10$^{-20}$ cm$^2$) | 0.773 | | 0.477 | 0.444 |
| δ$_{rel}$ (%) | 9.09 | | 5.90 | 5.50 |

Table 10. Adjustment parameters found with the different « modified » (KM) and « extended » (EXT. 1 and EXT. 2) J.O. approaches

Such a situation is not encountered in the case of the « Extended » J.O. approaches. This is illustrated in Fig. 6 which shows the variation of the relative standard deviation δ$_{rel}$ with the E(4f5d) energy. From these graphs, it is clear that a minimum RMS value is obtained for E(4f5d) ≈ 34 000 cm$^{-1}$ in the case of the EXT. 1 approach (« flat » 4f5d excited configuration) and for E(4f5d) ≈ 32 000 cm$^{-1}$ in the case of the EXT. 2 approach (full free-ion 4f5d excited configuration shifted down to the considered energy). We recall here that the first approach corresponds to the first approach successfully applied and published [12] in the case of the Nd$^{3+}$, Eu$^{3+}$ et Er$^{3+}$ ions.

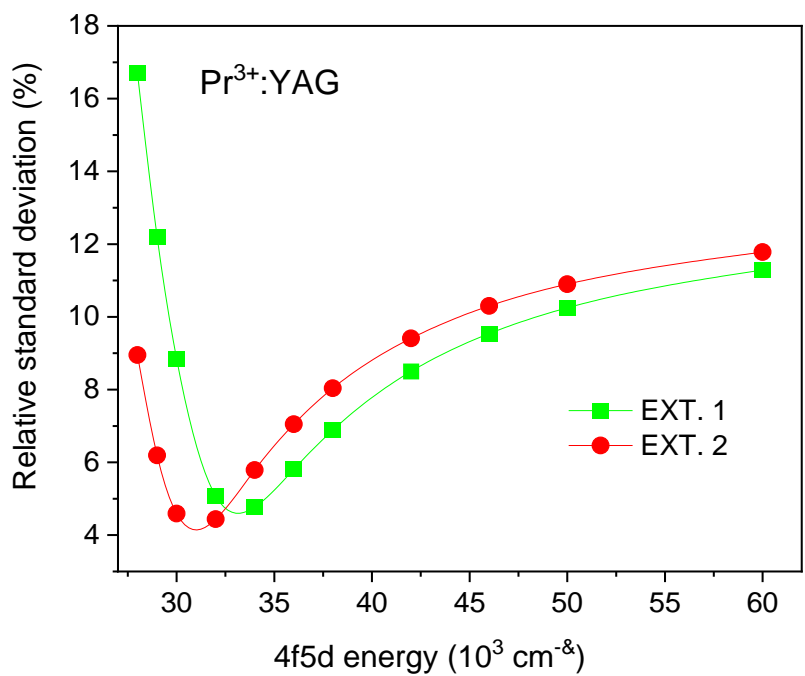

Figure 6 : Relative standard deviations obtained between experimental and calculated transition (or oscillator) strengths in the case of Pr$^{3+}$:YAG, by using the two « extended » EXT. 1 and EXT. 2 J.O. approaches, versus the 4f5d energy (lowest energy of the 4f5d excited configuration in the case of EXT. 2)

The figure 7 shows, for each considered absorption transition (from the $^3H_4$ ground state), the ratio between calculated and experimentally measured oscillator (or transition) strengths. The figure displays what is obtained by using the KM and the two EXT. 1 and EXT. 2 methods presented above.

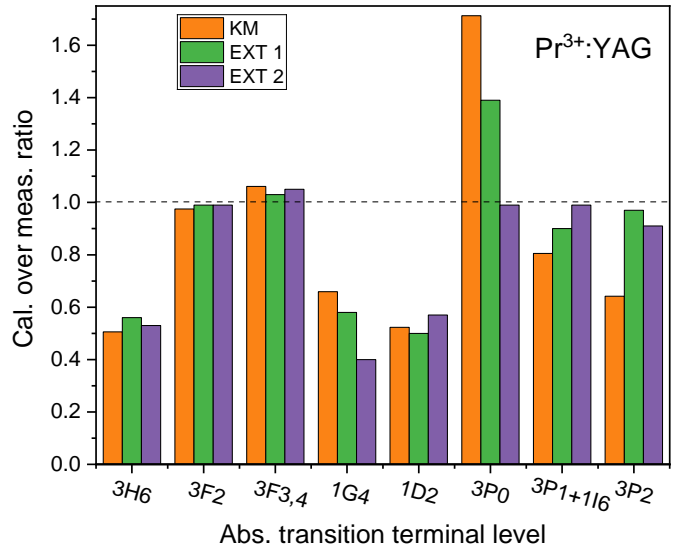

Figure 7 : Ratio between calculated and experimentally measured oscillator (or transition) strengths according to the different approaches (« modified » KM and « extended » EXT. 1 and EXT. 2 J.O. approaches)

The results are about the same for the absorption transitions to levels $^3H_6$, $^3F_2$, $^3F_{3,4}$ and $^1D_2$ with an excellent agreement between calculated and measured values for the transitions to $^3F_2$ and $^3F_{3,4}$. The results are more satisfactory with the « extended » approaches than with the « modified » one in the case of the absorption transitions to the important levels $^3P_2$, $^3P_0$ and $^3P_1+^1I_6$.

For the sake of clarity, we also gather in Table 11 all the radiative emission lifetimes calculated within the framework of the different theoretical approaches and a comparison is made with the effective fluorescence lifetimes measured between 77 and 300K. In the case of the $^3P_0$ emitting level, it is reported both the radiative emission lifetime calculated for the only $^3P_0$ level and the effective one calculated by accounting for the thermalized $^3P_{0,1}$ and $^1I_6$ levels. All the approaches lead to satisfactory results with slightly better values with the EXT. 1 « extended » one.

| Calculated radiative lifetimes (µs) | JO modified (KM) approach (with E4f5d = 34000 cm$^{-1}$) | JO Extended theory (EXT. 1) (flat with E4f5d = 34 000 cm$^{-1}$) | JO Extended theory (EXT. 2) (full shifted down to 32 000 cm$^{-1}$) | Measured fluorescence lifetimes (µs) |
|---|---|---|---|---|
| $\tau_R$, $\tau_{Reff}$ ($^3P_0$) | 9.91, 14.8 | 9.26, 13.4 | 12.6, 18.8 | 8.8-11.9 |
| $\tau_R$ ($^1D_2$) | 175 | 173 | 178 | 196-238 |
| $\tau_R$ ($^1G_4$) | 1160 | 1170 | 1520 | - |

Table 11. Calculated (radiative) and measured (fluorescence) lifetimes in Pr$^{3+}$:YAG with, In the last column, the two numbers corresponding to the radiative lifetimes measured at 300 and 77 K respectively.

Finally, we summarize the results concerning the branching ratios with the graphs reported in Fig. 8 and with the RMS standard deviations $\Delta\beta_{3P0}$, $\Delta\beta_{3P0}^{eff}$ and $\Delta\beta_{1D2}$ for the emissions coming from the $^3P_0$, $^3P_{0,1}+^1I_6$ and $^1D_2$ emitting levels in Table 12.

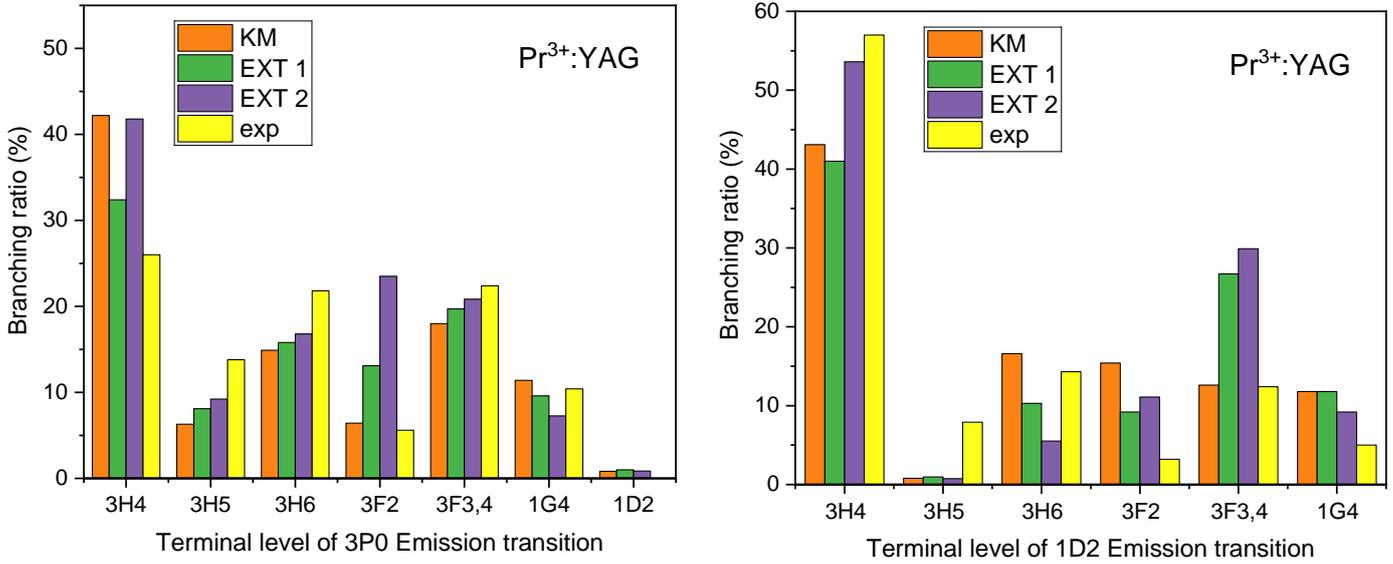

Figure 8 : Comparison of calculated and experimental branching ratios of the $^3P_{0,1}+^1I_6 \to A$ and $^1D_2 \to B$ emission transitions with A = $^3H_4, ^3H_5, ^3H_6, ^3F_2, ^3F_{3,4}, ^1G_4, ^1D_2$ and B = $^3H_4, ^3H_5, ^3H_6, ^3F_2, ^3F_{3,4}, ^1G_4$

| RMS deviations of branching ratios | JO modified (KM) approach (with E4f5d = 34000 cm$^{-1}$) | JO Extended theory (EXT. 1) with E4f5d = 34000 cm$^{-1}$ | JO Extended theory (EXT. 2) with full 4f5d shifted down to 32000 cm$^{-1}$ |
|---|---|---|---|
| $\Delta\beta_{3P0}$, $\Delta\beta_{3P0}^{eff}$ (%) | 14.5, 8.0 | 11.5, 5.4 | 13.5, 10.2 |
| $\Delta\beta_{1D2}$ (%) | 10.2 | 10.0 | 9.4 |

Table 12. RMS standard deviations $\Delta\beta_{3P0}$, $\Delta\beta_{3P0}^{eff}$ and $\Delta\beta_{1D2}$ for the emissions coming from the $^3P_0$, $^3P_{0,1}+^1I_6$ and $^1D_2$ emitting levels

According to these results, the best agreements between calculated and experimentally measured branching ratios for both $^3P_{0,1}+^1I_6$ and $^1D_2$ emitting levels are obtained with the EXT.1 « extended » J.O. approach.

### 3.5. Comparison with results obtained for Pr$^{3+}$:ZBLAN

To give an idea of what can be expected with a Pr$^{3+}$ doped material characterized by a weak (instead of a strong) local crystal field, thus by a low energy and probably less influential 4f5d excited electronic configuration, we report in the following figures and tables the results which we obtained in the case of the well-known Pr$^{3+}$-doped ZBLAN fluoride glass, a reference laser material for which complete luminescence data (oscillator and transition strengths, calibrated emission spectra and experimentally measured branching ratios) can be found in the literature [38, 39]. Details of the calculations are gathered in the **Supplementary Material.**

The Figure 9 presents first for each absorption transition (like in Figure 7 for Pr$^{3+}$:YAG) the ratio between calculated and experimentally measured oscillator (or transition) strengths obtained by using the standard J.O. approach (since it was not necessary to use a « modified » one to obtain a positive $\Omega_2$ parameter and a good RMS value) and the three « extended » ones noted EXT. 0, EXT. 1 and EXT. 2. EXT. 0 corresponds to a calculation based on the consideration of the full free-ion 4f5d excited electronic configuration (as it is determined by using the Cowan codes), EXT. 1 to a calculation based on a flat 4f5d configuration at the energy E(4f5d) = 38 000 cm$^{-1}$, and EXT. 2 to a calculation based on a full free-ion 4f5d configuration shifted down to the low energy in order to have the lowest levels at 36 000 cm$^{-1}$, both energies (38 000 cm$^{-1}$ and 36 000 cm$^{-1}$) corresponding (like in Fig. 6 for Pr$^{3+}$:YAG) to minimum RMS values in the two situations. Both energies also coincide with the position (maximum at about 250 nm and onset at about 270 nm, thus 40 000 cm$^{-1}$ and 37 000 cm$^{-1}$, respectively) of the small absorption bump (probably low energy levels of the first excited 4f5d configuration) observed in the near-UV absorption spectrum of Pr :ZBLAN [39, 48].

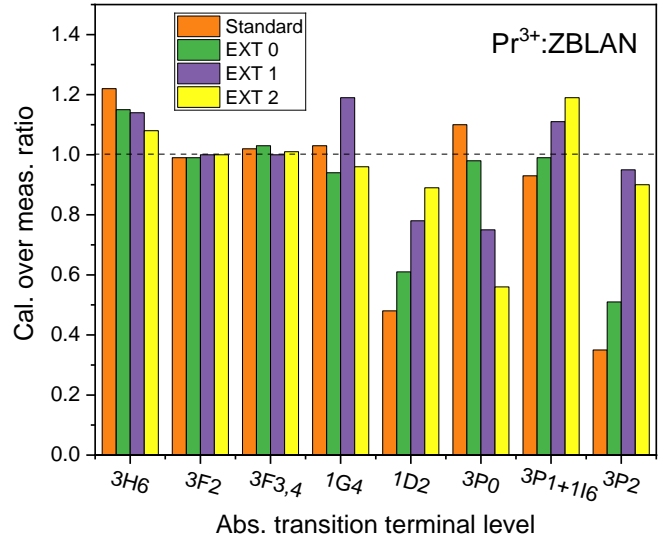

Figure 9. Ratio between calculated and experimentally measured oscillator (or transition) strengths according to the different approaches (« Standard » and « Extended » EXT. 0, EXT. 1 and EXT. 2 J.O. approaches)

As in case of Pr$^{3+}$:YAG (see in Fig. 7) the results are about the same for the absorption transitions to levels $^3H_6$, $^3F_2$ and $^3F_{3,4}$. However, the agreement between calculated and measured values for the transitions to $^3H_6$ and $^1G_4$ is much more satisfactory than in the case of Pr$^{3+}$:YAG. Concerning the transitions to the $^1D_2$ and $^3P_0$ levels, the best agreement is obtained with the EXT. 1 and EXT. 2 « extended » approaches for the former and with the EXT. 0 and the « standard » approaches for the latter. Concerning the transitions to the $^3P_1+^1I_6$ and $^3P_2$ levels, it is the reverse situation. Nevertheless, on the average (see in Table 13), the best agreement is obtained, like in the case of Pr$^{3+}$:YAG, by using the EXT. 1 « extended » approach.

| Parameters (10$^{-20}$ cm$^2$) | JO Standard approach | Parameters (a.u.) | JO Extended theory (EXT. 0) (full 4f5d) | JO Extended theory (EXT. 1) (flat with E4f5d = 38000 cm$^{-1}$) | JO Extended theory (EXT. 2) (full 4f5d shifted down to 36000 cm$^{-1}$) |
|---|---|---|---|---|---|
| $\Omega_2$ | 2.059 | $X_1$ | 1.052×10$^{-4}$ | 5.357×10$^{-5}$ | 5.730×10$^{-5}$ |
| $\Omega_4$ | 4.901 | $X_3$ | 4.728×10$^{-6}$ | 7.286×10$^{-7}$ | 5.513×10$^{-7}$ |
| $\Omega_6$ | 5.087 | $X_5$ | 1.007×10$^{-7}$ | 3.388×10$^{-8}$ | 3.697×10$^{-8}$ |
| $\delta$ (10$^{-20}$ cm$^2$) | 0.76 | | 0.58 | 0.14 | 0.22 |
| $\delta_{rel}$ (%) | 9.5 | | 7.23 | 1.81 | 2.76 |

Table 13. Adjustment parameters found with the different « standard » and « extended » (EXT. 0, EXT. 1 and EXT. 2) J.O. methods in the case of Pr$^{3+}$:ZBLAN

Finally, as above, we summarize the results concerning the emission transitions coming from the $^3P_{0,1}$+$^1I_6$ and $^1D_2$ emitting levels with the RMS standard deviations of the branching ratios $\Delta\beta_{3P0}^{eff}$ and $\Delta\beta_{1D2}$ reported in Table 14, and with the calculated and measured emission lifetimes reported in Table 15.

| RMS deviations of branching ratios | JO Standard approach | JO Extended theory (EXT. 0) (full 4f5d) | JO Extended theory (EXT. 1) (with E4f5d = 38000 cm$^{-1}$) | JO Extended theory (EXT. 2) (full 4f5d shifted down to 36000 cm$^{-1}$) |
|---|---|---|---|---|
| $\Delta\beta_{3P0}^{eff}$ (%) | 12.7 | 3.9 | 14.0 | 15.5 |
| $\Delta\beta_{1D2}$ (%) | 6.4 | 7.2 | 11.6 | 7.7 |

Table 14. RMS standard deviations $\Delta\beta_{3P0}^{eff}$ and $\Delta\beta_{1D2}$ for the emissions coming from the thermalized $^3P_{0,1}+^1I_6$ and the $^1D_2$ emitting levels

| Calculated radiative lifetimes (µs) | JO Standard | JO Extended theory (EXT. 0) (full free-ion 4f5d) | JO Extended theory (EXT. 1) (with E4f5d = 38000 cm$^{-1}$) | JO Extended theory (EXT. 2) (full 4f5d shifted down to 36000 cm$^{-1}$) | Measured fluorescence lifetimes (µs) [39] |
|---|---|---|---|---|---|
| $\tau_{Reff}$ (3P0) | 51.8 | 46 | 26.5 | 34.9 | 42 |
| $\tau_R$ (1D2) | 484 | 345 | 207 | 241 | 360 |
| $\tau_R$ (1G4) | 3.07 | 2.58 | 2.03 | 2.6 | - |

Table 15. Calculated (radiative) and measured emission lifetimes in Pr$^{3+}$:ZBLAN

According to these results, the best overall agreements between calculated and measured branching ratios and between calculated (radiative) and measured emission lifetimes are obtained by using the EXT. 0 « extended » J.O. approach.

## 4. Discussion on laser emission properties

The emission and gain cross section spectra associated with the $^3P_{0,1} \rightarrow {}^3H_4$ emission transition around 490 nm were already reported above in Fig. 5. The emission cross section spectra associated with the other main emission transitions coming from the thermalized $^3P_{0,1}$ and $^1I_6$ emitting levels are reported in Fig. 10. These spectra are obtained by using the intensity spectrum reported in Fig. 2 and the Fuchtbauer-Ladenburg (FL) expression (2) with the experimental branching ratios $\beta_{exp}^{eff}$ reported in Table 2 and the emission lifetime $\tau_{Reff} \approx 10$ μs. After examination of these spectra several remarks can be made.

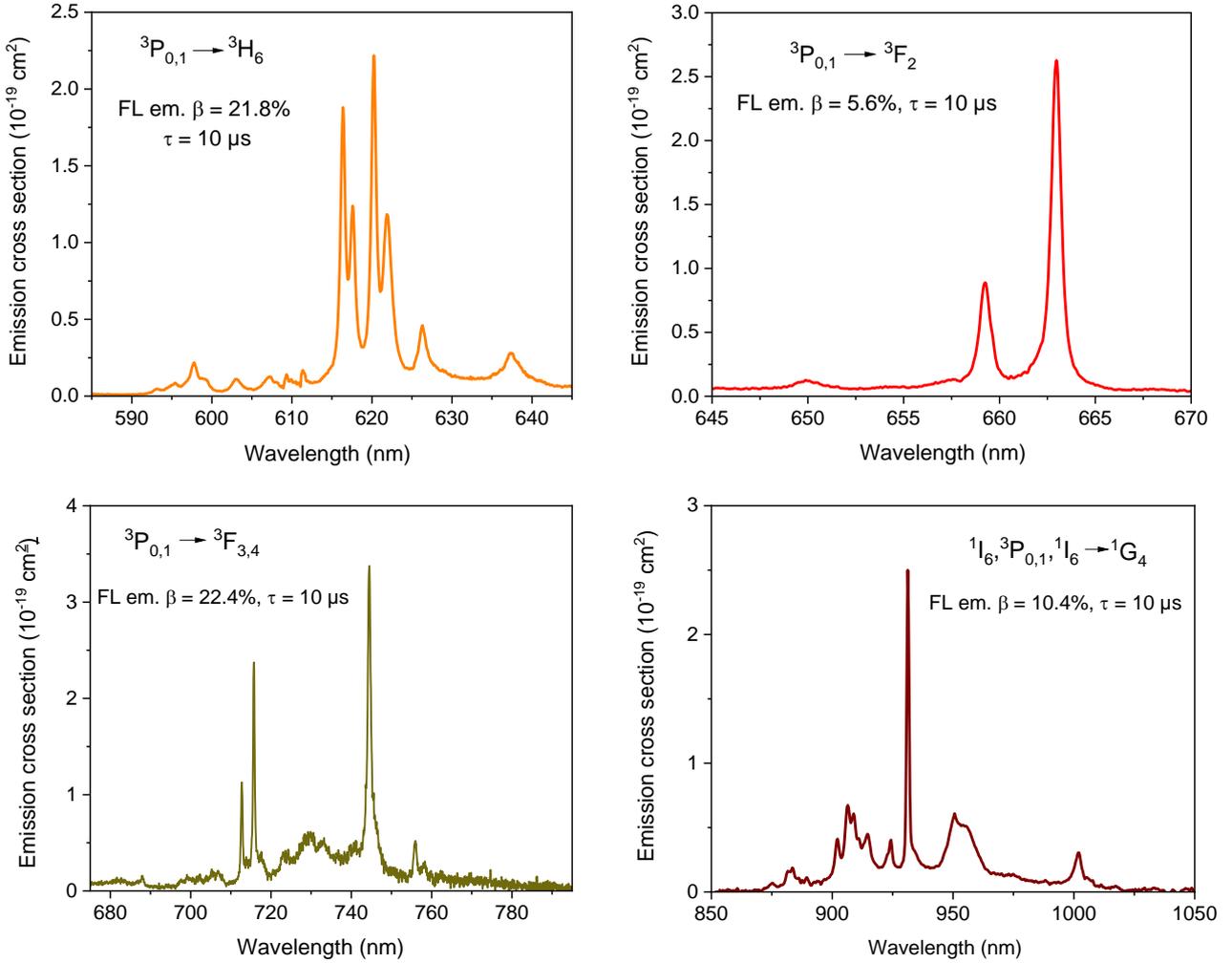

Figure 10 : Emission cross section spectra associated with the main emission transitions originating from the thermalized $^3P_{0,1}$ and $^1I_6$ energy levels.

1. All the emission transitions exhibit large gain cross sections, well exceeding $1.5 \times 10^{-19}$ cm². These cross sections are larger than that found in the reference laser fluoride crystal $Pr^{3+}$:LiYF$_4$ [49, 50], especially in the orange region around 620 nm and in the deep-red around 716 and 744.5 nm [49], which is due to sharper emission lines and a shorter effective emission lifetime (10 μs for $Pr^{3+}$:YAG against 50 μs for $Pr^{3+}$:LiYF$_4$).

2. Because of larger crystal-field splittings of energy levels allowing for detrimental excited-state absorption processes, very few $Pr^{3+}$-doped oxide crystals (none oxide glasses) led so far to noticeable laser emission properties. Among them, the most interesting crystal is the $Pr^{3+}$-doped perovskite YAlO$_3$ (YAP) [27-30] and, to a lesser extent, the stoichiometric phosphate LiPrP$_4$O$_{12}$ [31], the $Pr^{3+}$-doped strontium aluminates Sr$_{0.5}$La$_{0.5}$Mg$_{0.5}$Al$_{11.5}$O$_{19}$ [32], Sr$_{0.7}$La$_{0.3}$Mg$_{0.3}$Al$_{11.7}$O$_{19}$ (ASL) [33] and (Sr or Ca,Mg)Al$_{12}$O$_{19}$ [34-37] and the $Pr^{3+}$-doped garnet Y$_3$Al$_5$O$_{12}$ (YAG) [23, 24], the perovskite being characterized by lower phonon frequencies (585

cm$^{-1}$) [51] and lower crystal-field splittings [52] than the other compounds. In fact, only Pr$^{3+}$:YAP was lased at room temperature (under CW diode laser pumping) at green, red and deep-red emission wavelengths ([30] and refs therein). Pr$^{3+}$-doped YAG, on its side, could be also lased at room temperature, under flashlamp pumping [23], at the deep-red emission wavelength of 744.7 nm ($^3P_0 \to {}^3F_4$ transition, see in Fig. 9), and, under pulsed laser or CW laser pumping, at the green and orange emission wavelengths of 488 nm ($^3P_0 \to {}^3F_4$ transition) and 616 nm ($^3P_0 \to {}^3H_6$ transition), but only at low temperatures [24, 25], i.e. up to 32 K for the green emission [24] and up to 80 K for the orange one [24, 25]. It is worth noting here that lasing at the blue and green laser wavelengths was obtained without any cavity mirrors, only between the faces of the 0.6%Pr:YAG crystal used in the experiments [24], and that no attempts was ever made to lase in the red around 663 nm ($^3P_0 \to {}^3F_2$ transition) or in the near-infrared around 931 nm ($^3P_0,{}^1I_6 \to {}^1G_4$ emission transition).

3. According to the excitation spectra reported by Zhou et al [20], the two lower energy 4f5d absorption bands which characterize Pr$^{3+}$:YAG at room temperature should be peaking around 293 nm and 245 nm, the first one (which will be labelled 5d1) extending between 262 and 315 nm (31276-38170 cm$^{-1}$) and the second one (5d2) between 230 and 260 nm (38460-43480 cm$^{-1}$). Therefore, using the positions of the 4f$^2$ energy levels of Pr$^{3+}$:YAG [14, 15] which give rise to the absorption spectrum reported in Fig. 1 allows us to built the energy level diagram reported in Fig. 11.

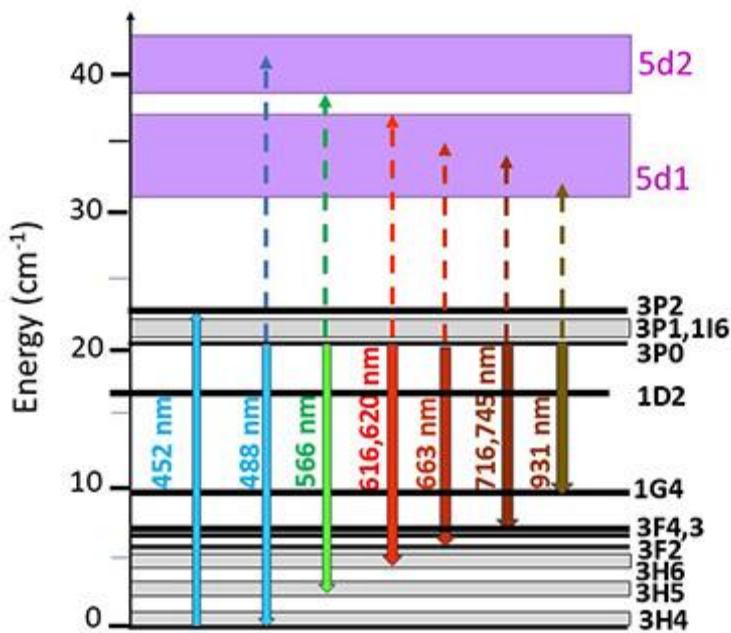

Figure 11 : Energy level scheme and main absorption, emission (full arrows) and excited-state absorption (dashed arrows) transitions occurring in Pr$^{3+}$:YAG

According to this diagram, excited-state absorption (ESA) at the green and deep-red laser wavelengths of 488 nm and 744.7 nm mentioned above should fall right into the middle, thus at the maxima, of the two (5d1 and 5d2) 4f5d absorption bands. ESA at the red one of 616 nm should be less detrimental. Therefore, lasing at 488 nm at very low temperature (below 32 K) and 744.7 nm at room temperature likely means that gain cross sections at these laser wavelengths are larger than ESA, at least at the respective temperatures. According to Fig. 5 and Fig. 9, this gives an upper limit to ESA cross sections of about 3×10$^{-19}$ cm$^2$. Therefore, lasing should be possible, after blue diode pumping around 452 nm ($^3H_4 \to {}^3P_2$ absorption), at room temperature, and both between the two 4f5d absorption bands and below them, i.e. at the green and near-infrared wavelengths of 566 nm (see in Fig. 5) and 931 nm (see in Fig. 10).

## 5. Conclusions and future works

The first part of the article consisted in revisiting the luminescent properties $Pr^{3+}$:YAG, by improving and completing the spectroscopic data which could be found in the past literature. Absorption spectra were registered first in a more extended wavelength domain and more reliable absorption line intensities (oscillator and transition strengths) with improved inter-multiplet transition assignments could be reported. Emission spectra were also registered, and carefully corrected for the spectral response of the detection system in the whole visible and near-infrared spectral domain, allowing for the determination of more complete and more reliable emission transition branching ratios. Some fluorescence decay measurements were also performed at low and high temperatures to corroborate the values of emission lifetimes which were reported in the past by using different types of samples with different $Pr^{3+}$ ion concentrations.

These experimental data were then used to show the results which can be obtained by using usual « Standard » and « Modified » versions of the Judd-Ofelt formalism and to show the improvements which can be found by using an extension of this theory recently applied to the cases of $Eu^{3+}$, $Nd^{3+}$ and $Er^{3+}$ doped materials with success [11, 12]. For that purpose, three types of « Extended » approaches labelled EXT. 0, EXT. 1 and EXT. 2 were considered. EXT. 0 consists in considering that the electric-dipole character of each absorption transition is enabled due to the contribution of various levels of the excited 4f5d electronic configuration, as they are determined for the free ion using the well-known Cowan computer codes. EXT. 2 follows about the same procedure but by assuming that all the free-ion energy levels can be shifted by a chosen quantity. EXT. 1 corresponds to a situation close to that used in the « Standard » and « Modified » methods. It consists in assuming that all the levels of the 4f5d configuration are degenerate (flat configuration) and fixed at a given (tunable) energy.

The results show that the "Standard" and, actually, the "EXT. 0" J.O. approaches cannot be used in the case of $Pr^{3+}$:YAG, since they both lead to non-physical negative $\Omega_2$ and $X_1$ parameters, respectively. Reasonable results are obtained with the "Modified" method but with 4 adjustable parameters $\Omega_2$, $\Omega_4$, $\Omega_6$ and $\alpha$, the latter being chosen to match the most significant 4f5d energetic position i.e. the position of the lowest 4f5d absorption band observed around 34 000 cm$^{-1}$ in the case of $Pr^{3+}$:YAG. This value was not chosen to get the best agreement between measured and calculated absorption line strengths, such a best agreement being obtained for non-realistic 4f5d energy values, but to lead to the most reasonable $^3P_0$ and $^1D_2$ emission lifetimes. Better results are obtained with the two other "Extended" J.O. approaches, especially with "EXT. 2" for the description of the absorption line strengths, but also with "EXT. 1" for the description of the $^3P_0$ and $^1D_2$ emission lifetimes and emission transition branching ratios. Fitting of the data thus could be performed with the aid of three odd and positive parameters $X_1$, $X_2$ and $X_3$ and by choosing the 4f5d energy (that corresponding to the flat configuration considered in the case of EXT. 1 or that corresponding to the energy down to which the full free-ion 4f5d configuration is shifted in the case of EXT. 2) leading to the best agreement between measured and calculated absorption line strengths, the minima being obtained for energies of 34 000 cm$^{-1}$ (as in the case of the "Modified" J.O. approach) and 32 000 cm$^{-1}$. More particularly, whereas all the methods (except the "Standard" and "EXT. 0) lead to the same results for the absorption transitions to levels $^3H_6$, $^3F_2$, $^3F_{3,4}$ and $^1D_2$, with an excellent agreement between calculated and measured values for the transitions to $^3F_2$ and $^3F_{3,4}$, the results are more satisfactory with the EXT. 1 and EXT. 2 « Extended » approaches than with the « Modified » one in the case of the absorption transitions to the important levels $^3P_2$, $^3P_0$ and $^3P_1+^1I_6$. Moreover, what is not allowed within the framework of the « Standard » and « Modified » approaches, the « Extended » ones clearly show, as it is observed, the possibility of emission transitions from the $^3P_0$ emitting level down to levels with odd J quantum numbers such as the $^3P_0 \rightarrow ^3H_5$ (around 560 nm) and $^3P_0 \rightarrow ^3F_3$ (around 730 nm) with non-negligible (although still not high enough) branching ratios. All in all, the "EXT. 1" gives the most satisfactory results.

Parallelly (what is more particularly developed in the « Supplemental ») a comparison is made by examining the results which can be obtained with a $Pr^{3+}$-doped material characterized by a higher enegy 4f5d excited configuration than in the case of $Pr^{3+}$:YAG. Choice was made of the well-known $Pr^{3+}$:ZBLAN fluoride glass and the results were the following ones. Unlike the case of $Pr^{3+}$:YAG, reasonably good and physically significant results (with positive fitting parameters) could be obtained by using the « Standard » and « EXT. 0 » J.O. approaches, without resorting to a « Modified » one. Moreover, better results in terms of line strengths, emission lifetimes and branching ratios could be obtained with the « EXT. 0 » approach than with the « Standard » and the other « EXT. 1 » and « EXT. 2 » extended approaches. This proves a lower influence of the 4f5d excited configuration on the spectroscopic properties of $Pr^{3+}$:ZBLAN than in the case of $Pr^{3+}$:YAG, and, at the same time, a significant improvement of the results when the full 4f5d free-ion configuration is included in the calculations.

The above revisited spectroscopic and luminescence data were finally used to address the questions of $Pr^{3+}$:YAG emission gain cross sections and laser operation at the various emission wavelengths which characterize the $^3P_0$ emitting level. Thanks to the registration of well-calibrated emission spectra and the estimation of reliable branching ratios, it is shown that laser operation could be achieved at several emission wavelengths in the visible and near-infrared spectral domains, especially around 566 nm and 931 nm, by using adequate laser cavities and laser mirrors (not between the faces of the crystal, as it was demonstrated in the past) and by pumping the laser crystal with a blue laser diode at about 452 nm or an optically-pumped and frequency-doubled semiconductor laser (OPSL) at about 488 nm, as it was realized with other $Pr^{3+}$-doped laser crystals and glasses.

**Acknowledgements**

Calculations have been performed using HPC resources from the DNUM CCUB (Centre de Calcul de l'Université de Bourgogne). Thanks are expressed to Dr. M. Malinowski who provided in the past the $Pr^{3+}$:doped YAG crystals used in this study.